\documentclass[a4paper,12pt,oneside,onecolumn]{elsarticle}
\usepackage{tabularx} 
\usepackage{amsmath}  
\usepackage{amsthm}
\usepackage{graphicx} 
\usepackage[margin=1in,letterpaper]{geometry} 

\usepackage{xcolor}
\usepackage[final]{hyperref} 
\hypersetup{
	colorlinks=true,       
	linkcolor=blue,        
	citecolor=blue,        
	filecolor=magenta,     
	urlcolor=blue         
}

\usepackage{amssymb}
\usepackage{caption}
\usepackage{subcaption}
\usepackage{tikz}
\usepackage{soul}
\usepackage{slashed}
\usetikzlibrary{patterns}
\usetikzlibrary{shapes.arrows}
\usepackage{pgfplots}
\usepackage{placeins}
\pgfplotsset{compat=newest}
\usepackage{mathtools} 
\usepackage{diagbox}
\usepackage{fixltx2e}
\usetikzlibrary{arrows.meta}
\usetikzlibrary{calc,decorations.markings,fit,shapes,chains,arrows}
\usetikzlibrary{fillbetween,backgrounds}
\usetikzlibrary{patterns}
\usepackage{lineno}
\usepackage{adjustbox}
\usepackage{booktabs}%
\usepackage{algorithm}
\usepackage{algorithmic}
\newtheorem{remark}{Remark}

\begin{document}

\begin{frontmatter}

\title{An Efficient Phase-field Framework for Contact Dynamics between Deformable Solids in Fluid Flow}
\author[ubc]{Biswajeet Rath}
\ead{biswajee@mail.ubc.ca}

\author[ubc]{Xiaoyu Mao}
\ead{dgcsmaoxiaoyu@gmail.com}

\author[ubc]{Rajeev K. Jaiman\corref{cor1}}
\ead{rjaiman@mail.ubc.ca}
\cortext[cor1]{Corresponding author}
\address[ubc]{Department of Mechanical Engineering, The University of British Columbia, Vancouver, BC V6T 1Z4}

\begin{abstract}
    Elastic contact in hydrodynamic environments is a complex multiphysics phenomenon and can be found in applications ranging from engineering to biological systems. Understanding the intricacies of this coupled problem requires the development of a generalized framework capable of handling topological changes and transitioning implicitly from fluid-structure interaction (FSI) conditions to solid-solid contact conditions. We propose a mono-field interface advancing method for handling multibody contact simulations in submerged environments. Given the physical demands of the problem, we adopt a phase-field based fully Eulerian approach to resolve the multiphase and multibody interactions in the system. We employ a stabilized finite element formulation and a partitioned iterative procedure to solve the unified momentum equation comprising solid and fluid dynamics coupled with the Allen-Cahn phase-field equation. The evolution of solid strain in the Eulerian reference frame is governed by the transport of the left Cauchy Green deformation tensor. We introduce a contact force approach to handle smooth elastic-elastic and elastic-rigid contact based on the overlap of the diffused interfaces of two colliding bodies. We propose a novel approach to extend the model for multibody contact simulations while using a single phase-field function for all the solids. The method is based on updating the solid boundaries at every time step and checking for collisions among them. The developed approach eliminates the need to solve multiple phase field equations and multiple strain equations at every time step. This reduces the overall computational time by nearly $16\%$ compared to a multi phase-field approach. The implemented model is verified for smooth dry contact and FSI contact scenarios. Using the proposed framework, we demonstrate the collision dynamics between multiple bodies submerged in an open liquid tank.

\smallskip
\smallskip


\end{abstract}
\end{frontmatter}


\section{Introduction} \label{sec1}
Fluid-Structure interaction (FSI) is an omnipresent and complicated phenomenon that manifests itself in natural, biological and engineering systems. The relative motion between a solid and the fluid in both internal and external flows leads to the application of loads on the solid, resulting in deformations and/or displacements in it. 
Our primary motivation comes from ship-ice interaction in Arctic regions during the collision of ships with large ice sheets.  
The framework should be able to identify different types of interfaces and handle various structural-level interactions apart from free-surface motion and fluid-structure interaction. Simulating the process of ship-ice interactions involves handling drifting, large deformation, collisions, fracture and fragmentation of the ice sheets \cite{lu2018large, li2020finite, shi2023numerical}. 
In such scenarios, accurately capturing contact dynamics with hydrodynamic interactions poses significant challenges for computational modeling. Typically, solid-solid contact problems are predominantly modeled within a Lagrangian framework, where the discretization moves with the deforming objects. By considering an Eulerian formulation, our aim is to develop an integrated and versatile model capable of effectively handling contact dynamics with fluid-structure interaction. 

In Lagrangian formulations, numerical modeling of contact is a challenging task for two important reasons. First, both the contact boundary and the contact boundary condition i.e. contact force are unknown \textit{a priori} and are a part of the solution. The second reason is the discontinuous nature of the contact force which introduces an inherent irregularity and non-smoothness in the problem. Mathematically, these issues translate into a set of constraints known as Hertz-Signorini-Moreau (HSM) conditions \cite{kikuchi1988contact}. They consist of a non-penetration constraint and a non-zero contact force condition. This results in a set of variational inequalities for the overall problem \cite{kikuchi1988contact, wriggers2006computational}. One of the earliest approaches for modeling contact for large deformation bodies involving non-matching meshes in a Lagrangian frame of reference was the node-to-surface algorithm introduced by Hughes et al. \cite{hughes1976finite} and improved in some of the later works in \cite{wriggers1990finite, taylor1991patch}. The method is based on the idea of checking for contact between a node on the slave body and a surface element on the master body. The contact constraints are then imposed onto the set of slave nodes (also known as active set) based on the type of solution method i.e. interior point method, Lagrange multiplier method, penalty method, etc. The choice of an appropriate slave body and concentrated nodal forces were some of the limitations of the approach. 

Within the Lagrangian formalism, several weak (integral) implementations of the contact constraints have been suggested to bypass the drawbacks of the node-to-surface algorithm. The most significant among them was the surface-to-surface contact approach \cite{taylor1991patch, zavarise1998segment}. The contact contribution to the weak form is integrated by a set of \textit{a priori} defined Gauss quadrature points on each element of the slave surface. This particular algorithm also suffers from the master/ slave bias similar to the node-to-surface method.
However, the improved two-pass versions of the above algorithms, which treat both bodies as slaves in two alternate iterations, remove this inherent bias \cite{el2001stability, zavarise2011contact}. Mortar methods \cite{belgacem1998mortar, mcdevitt2000mortar, farah2015segment} are another class of approaches for resolving contact in a weak sense as they allow for a variationally consistent treatment of non-penetration and frictional sliding conditions. 
Recently, some of the above methods have been implemented in isogeometric analysis (IGA) formulations \cite{hughes2005isogeometric, cottrell2009isogeometric} for their improved accuracy and continuity properties by considering the same basis functions for geometry and its discretization to enforce contact constraints. The authors in \cite{de2014isogeometric, fahrendorf2022isogeometric, temizer2011contact, otto2020explicit} illustrate the applicability of IGA to problems involving contact mechanics and impact dynamics. 

In the context of FSI applications, we have the additional traction continuity condition at the fluid-solid interface. Thus, we need to ensure a smooth transition between the FSI conditions and the contact conditions over the solid boundary. 
The presence of the fluid and the requirement to handle large solid deformations make the Eulerian frame of reference more suitable for our applications.
Keeping in mind the physical requirements of the problem, we utilize a fully Eulerian-based FSI framework for the present work, 
whereby the space itself is discretized and the evolving solid bodies are transported through it. A fully Eulerian approach alleviates the issues of dissimilar kinematics and dynamics between the fluid and solid domains by formulating the governing conservation laws in a unified framework. 
The Eulerian approach can significantly reduce the complexity of contact detection and the satisfaction of contact constraints.
Some of the recent studies have demonstrated the simplicity of handling solid-solid contact in both isolated \cite{levin2011eulerian, lorez2024eulerian} and submerged environments \cite{valkov2015eulerian, frei2016eulerian} within a fully Eulerian framework. 

One of the obvious ways to handle the non-penetration constraint in contact is to use optimization algorithms for solving the quadratic program. Such an approach was carried out in \cite{levin2011eulerian} where they used the Interior Point method in computer graphics simulations of colliding bodies. They utilized a finite difference fully Eulerian solver to carry out the structural simulations and used the reference map to evaluate the solid strains. The velocities of the solids were updated separately to keep track of the individual objects and a collision detection algorithm \cite{le2007broad} was used for broad-phase contact handling. However, using optimization algorithms is ill-suited and tedious for complex domain discretizations and diffuse interface surfaces. Hence, in a later work \cite{teng2016eulerian} by the above groups, they shifted to an explicit impulse-based approach for handling contact. 

An explicit contact force approach is the most common method for dealing with collisions in fully Eulerian FSI simulations.  
Such an explicit force can be represented in a generalized form as:
\begin{equation}
    \boldsymbol{f}_c = \kappa \psi(x) \boldsymbol{n},
\end{equation}
where $\psi(x)$ is the short-range influence function that is dependent on the gap between the bodies, $\kappa$ is a constant used to scale the force and $\boldsymbol{n}$ is the unit normal vector. 
Different forms of the influence function have been used in literature. \cite{valkov2015eulerian} used a smooth Heaviside function of the relative level set as their influence function to compute the contact force. The authors in \cite{burman2020nitsche, frei2016eulerian} formulated their influence function as inversely proportional to the gap between the colliding bodies. Such a form of the forcing function ensures that true contact never occurs and a fluid layer is always present between the colliding bodies. However, the contact force for such a model can have very high values and might cause stability issues in the numerical solver. On the other hand, \cite{lorez2024eulerian} proposed that the influence function by considering the overlap of different phase-field functions for their two body contact simulations. Such a model allowed for some inter-penetration between the bodies in contact.
In the present article, we introduce a phase-field based spring-type contact force as an additional body force that acts on the colliding bodies and prevents interpenetration.
  
Using a fully Eulerian framework, we adopt the phase-field model for the spatio-temporal evolution of the implicit interface. The phase-field interpolation and the unified formulation implicitly satisfy the velocity and stress continuity conditions at the fluid-solid interface. Although the phase-field based FSI formulation is not new and has been carried out previously in \cite{sun2014full, mokbel2018phase}, we focus exclusively on contact handling and propose a novel method to model multibody contact in submerged environments while using a single phase-field function for all the solids. We build upon our previous works \cite{mao2023interface, rath2023interface} of a stabilized finite element discretization for the fully Eulerian framework and implement a phase-field based contact force model to handle collisions between solids in a fluid environment. This alleviates the issue of a discontinuous contact force, since now the contact force can build up gradually as the diffuse boundaries of the colliding bodies overlap each other. We verify the implemented contact routine with the traditional Hertz contact model for smooth elastic solids. Such verification studies are important for ensuring the accuracy of the contact model before studying the FSI contact dynamics. We further verify our model for elastic-elastic contact in submerged environments. Finally, we present two application problems to demonstrate the robustness of the proposed multibody contact model. The presented method obviates the need for defining multiple phase-field functions and strain variables, thus saving significant computational resources otherwise required for solving the extra PDEs. 

The present article is structured as follows. In Section 2, we briefly review the fully Eulerian formulation and present the governing equations deriving from the conservation laws. We also discuss the evaluation of solid stresses in a fully Eulerian framework. In Section 3, we present the mono-field interface advancing (MFIA) method for collision handling. We elaborate on the MFIA algorithm and the contact force computations for the proposed formulation. Details of the numerical implementation are presented in Section 4, which emphasizes the temporal discretization and the system of matrix equations to be solved. Section 5 focuses on the verification and convergence studies of the implemented model for smooth dry contact and FSI contact scenarios. In Section 6, we present the results obtained from two applications to demonstrate the robustness of the proposed method. Section 7 summarizes the conclusions and provides directions for future work.

\section{Fully Eulerian Formulation} \label{sec2}
In this section, we provide a brief background of the fully Eulerian formulation. Both the solid and fluid equations are represented in Eulerian coordinates. From standard conventions in continuum mechanics, let $\Omega$ denote the domain in the reference configuration $\boldsymbol{X}$ at $t=0$ and $\Omega^t$ denote the deformed state of the domain in spatial coordinates, $\boldsymbol{x}$. The fluid and the solid make up the two parts of the domain, such that the total domain is given as $\Omega = \Omega_f^t \cup \Omega_s^t$ and the diffuse interface is defined as $\Gamma_{FSI}^t = \Omega_f^t \cap \Omega_s^t$. We consider both the solid and the fluid to be incompressible. We analyze the interactions between a Newtonian fluid and a hyperelastic solid for all our FSI cases.

\subsection{Continuum Equations}
We can express the unified continuum equations for the physical domains via a phase indicator $(\phi(\boldsymbol{x},t))$ based interpolation as follows
\begin{equation}\label{eq:UC}
	\begin{aligned}
		\nabla \cdot \boldsymbol{v} &= 0 \ \  &\text{on   } \Omega, \\
		\rho(\phi) \left(\frac{\partial \boldsymbol{v}}{\partial t}\bigg\rvert_x + (\boldsymbol{v} \cdot \nabla)\boldsymbol{v}\right) &= \nabla \cdot (\boldsymbol{\sigma}(\phi)) + \boldsymbol{b}(\phi) \ \  &\text{on   } \Omega,
	\end{aligned}
\end{equation}
where the physical properties are defined by the following interpolation rules:
\begin{equation}\label{eq:phase_int}
	\begin{aligned}
		\rho(\phi)&=\phi_f\rho_f + \phi_s\rho_s ,\\
		\mu(\phi)&=\phi_f\mu_f + \phi_s\mu_s .
	\end{aligned}
\end{equation}
In Eq.~(\ref{eq:UC}), $\boldsymbol{v}$ denotes the velocity and $\boldsymbol{b}$ denotes the body force at each spatial point $\boldsymbol{x} \in \Omega$. The first equation ensures mass conservation of the incompressible continuua in the domain in Eulerian coordinates. The second equation is the momentum balance equation for the combined fluid-solid domain. We vary the order parameter in the range $[-1,1]$ in the current work. Thus, we have $\phi_f = \frac{1-\phi}{2}$ and $\phi_s = \frac{1+\phi}{2}$ as a simple choice for the interpolation functions in Eq. (\ref{eq:phase_int}). We expand more on the phase indicator function in Section \ref{phase-field}. Interested readers are referred to our previous works in \cite{rath2023interface, mao2023interface} to review the process of arriving at the above set of equations in an Eulerian reference frame.

The semi-discrete variational form for the above differential equation, in conjunction with the generalized-alpha time integration method \cite{chung1993time, jansen2000generalized} can be presented as follows. Let $\mathcal{S}^h$ be the space of trial solutions whose values satisfy the Dirichlet boundary conditions and $\mathcal{V}^h$ be the space of test functions whose values vanish on the Dirichlet boundary. Thus, find $[\boldsymbol{v}_h^{n+\alpha}, p_h^{n+1}] \in \mathcal{S}^h$ such that $\forall [\boldsymbol{\psi}_h, q_h] \in \mathcal{V}^h$,
\begin{equation}\label{eq:var_UC}
	\begin{aligned}
		&\int_{\Omega} \rho(\phi)\left(\partial_{t} \boldsymbol{v}_{\mathrm{h}}^{\mathrm{n}+\alpha_{\mathrm{m}}}+(\boldsymbol{v}_{\mathrm{h}}^{\mathrm{n}+\alpha} \cdot \nabla) \boldsymbol{v}_{\mathrm{h}}^{\mathrm{n}+\alpha}\right) \cdot \boldsymbol{\psi}_{\mathrm{h}} \mathrm{d} \Omega+\int_{\Omega} \boldsymbol{\sigma}_{\mathrm{h}}^{\mathrm{n}+\alpha}: \nabla \boldsymbol{\psi}_{\mathrm{h}} \mathrm{d} \Omega +\int_{\Omega} q_{\mathrm{h}}\left(\nabla \cdot \boldsymbol{v}_{\mathrm{h}}^{\mathrm{n}+\alpha}\right) \mathrm{d} \Omega \\
		&+\sum_{\mathrm{e}=1}^{\mathrm{n}_{\mathrm{el}}} \int_{\Omega_{\mathrm{e}}} \frac{\tau_{\mathrm{m}}}{\rho(\phi)}\left(\rho(\phi) (\boldsymbol{v}_{\mathrm{h}}^{\mathrm{n}+\alpha} \cdot \nabla) \boldsymbol{\psi}_{\mathrm{h}}+\nabla q_{\mathrm{h}}\right) \cdot \boldsymbol{\mathcal{R}}_{\mathrm{m}}(\boldsymbol{v}, p) \mathrm{d} \Omega_{\mathrm{e}} +\sum_{\mathrm{e}=1}^{\mathrm{n}_{\mathrm{el}}} \int_{\Omega_{\mathrm{e}}} \nabla \cdot \boldsymbol{\psi}_{\mathrm{h}} \tau_{\mathrm{c}} \rho(\phi) \mathcal{R}_{\mathrm{c}}(\boldsymbol{v}) \mathrm{d} \Omega_{\mathrm{e}} \\
		&=\int_{\Omega} \boldsymbol{b}\left(t^{\mathrm{n}+\alpha}\right) \cdot \boldsymbol{\psi}_{\mathrm{h}} \mathrm{d} \Omega+\int_{\Gamma_{\mathrm{h}}} \boldsymbol{h} \cdot \boldsymbol{\psi}_{\mathrm{h}} \mathrm{d} \Gamma .
	\end{aligned}
\end{equation}
The first line consists of the Galerkin terms for the combined momentum and continuity equations. The second line contains the Petrov-Galerkin stabilization terms for the continuum equations. $\boldsymbol{\mathcal{R}}_m$ and $\mathcal{R}_c$ are the element-wise residuals for the momentum and continuity equations respectively. The stabilization parameters $\tau_m$ and $\tau_c$ \cite{shakib1991new, johnson2012numerical} are given by
\begin{equation}
	\tau_m = \left[\left(\frac{2}{\Delta t}\right)^2 + \boldsymbol{v}_h \cdot \boldsymbol{G} \boldsymbol{v}_h + C_I \bigg(\frac{\mu(\phi)}{\rho(\phi)}\bigg)^2 \boldsymbol{G}:\boldsymbol{G}\right]^{-1/2},  \text{   } \tau_c = \frac{1}{\mathrm{tr}(\boldsymbol{G})\tau_m} ,
\end{equation}
where $C_I$ is a constant derived from the element-wise inverse estimate and $\boldsymbol{G}$ is the element contravariant metric tensor. $n_{el}$ and $\Omega_e$ in Eq. \ref{eq:var_UC} denote the total number of elements and the volume occupied by each element, respectively. We elaborate on the strain evolution in the solid domain in the next sub-section.

\subsection{Solid Strain Evolution}
As we do not solve for the solid displacements in an Eulerian setting, we need an additional equation to evolve the solid strains and close the system of equations. This obviates the need for a mesh motion equation and helps us compute the stresses in the structures accurately. We capture the solid deformation by the evolution of the left Cauchy-Green deformation tensor ($\boldsymbol{B}$). Some of the other approaches used in the literature are the transport of the reference/ inverse map function ($\boldsymbol{\xi}$) \cite{valkov2015eulerian, rycroft2020reference} or the deformation gradient tensor ($\boldsymbol{F}$) \cite{liu2001eulerian, trangenstein1991higher}. Although $\boldsymbol{\xi}$ has the additional advantage of storing the location information, it can be inconvenient for actual stress computations, as it involves second-order derivatives. For the choice between $\boldsymbol{F}$ and $\boldsymbol{B}$, we go with $\boldsymbol{B}$ as it is a symmetric tensor and saves us the effort of computing four different components in 2-D or nine different components in 3-D. We emphasize the decomposition of the stress tensor into its volumetric and deviatoric components to ensure a uniform definition of pressure in the domain in Section \ref{sec2.3}.  

The transport equation for $\boldsymbol{B}$ in a solid moving with velocity $\boldsymbol{v}$ is given as
\begin{equation}
	\frac{\partial \boldsymbol{B}}{\partial t} + \phi_s (\boldsymbol{v} \cdot \nabla) \boldsymbol{B} = \phi_s \bigg((\nabla \boldsymbol{v}) \boldsymbol{B} + \boldsymbol{B}(\nabla \boldsymbol{v})^T \bigg) \ \ \ \  \text{on   } \Omega^s,
\end{equation}
where the pre-factor $\phi_s$ is given by $\phi_s=(\frac{1+\phi}{2})$.
Such an interpolation ensures that the above transport equation is solved only in the solid domain and smoothly extended into the interface. There is no physical significance in defining a strain measure in the fluid domain. To avoid any instabilities that might arise from the fluid side and creep up into the solid, we impose a Dirichlet condition of $\boldsymbol{B} = \boldsymbol{I}$ in the fluid domain at a cutoff for the phase-field function of $\phi<-0.95$ \cite{sugiyama2011full}. 

The weak form for the transport of the left Cauchy-Green deformation tensor is presented below. Let $\mathcal{S}^h$ be the space of trial solutions whose values satisfy the Dirichlet boundary conditions and $\mathcal{V}^h$ be the space of test functions whose values vanish on the Dirichlet boundary. The variational form of the left Cauchy-Green tensor equation can be written as: find $\boldsymbol{B}_h^{n+\alpha} \in \mathcal{S}^h$ such that $\forall \boldsymbol{m}_h \in \mathcal{V}^h$,
\begin{equation}
	\begin{aligned}
		&\int_{\Omega} (\boldsymbol{m}_h):\bigg( \partial_t \boldsymbol{B}_h^{\mathrm{n+\alpha_m}} + \phi_s (\boldsymbol{v}^{{\mathrm{n+\alpha}}} \cdot \nabla) \boldsymbol{B}_h^{\mathrm{n+\alpha}} \bigg) \mathrm{d} \Omega - \\
		&\int_{\Omega} \bigg( \phi_s \nabla \boldsymbol{v}^{{\mathrm{n+\alpha}}} \boldsymbol{B}_h^{\mathrm{n+\alpha}} + \phi_s \boldsymbol{B}_h^{\mathrm{n+\alpha}} (\nabla \boldsymbol{v}^{{\mathrm{n+\alpha}}} )^T \bigg) : (\boldsymbol{m}_h) \mathrm{d}\Omega \\
            &+\sum_{\mathrm{e}=1}^{\mathrm{n}_{\mathrm{el}}} \int_{\Omega_{e}} \tau_{\boldsymbol{B}}\left(\boldsymbol{v}^{\mathrm{n+\alpha}} \cdot \nabla \boldsymbol{m}_{\mathrm{h}}\right) : \boldsymbol{\mathcal{R}}(\boldsymbol{B}_{\mathrm{h}}) \mathrm{d} \Omega_{e} = 0,	
	\end{aligned}
\end{equation}
where $(:)$ denotes the double dot product and $\boldsymbol{\mathcal{R}}(\boldsymbol{B}_{\mathrm{h}})$ represents the element-wise residual of the strain evolution equation. The first and second lines contain the Galerkin projection terms and the third line contains the Petrov-Galerkin stabilization terms, with the stabilization parameter given by $\tau_{\boldsymbol{B}} = \bigg(\left( \frac{2}{\Delta t} \right)^2 + \boldsymbol{v}_h \cdot \boldsymbol{G} \boldsymbol{v}_h \bigg)^{-\frac{1}{2}}$. We obtain the updated values of $\boldsymbol{B}$ by solving the above variational equation in the solid domain which can then be used to calculate the new solid stresses before substituting in the unified continuum equations in the next time step.

\subsection{Stress Computation for Incompressible Materials} \label{sec2.3}
In this sub-section, we delve deeper into evaluating the stress for incompressible materials in a fully Eulerian framework. The total stress in the domain in Eq. \ref{eq:UC} can be expressed as a weighted average of the fluid and solid stresses based on the phase-field function:
\begin{equation}
    \boldsymbol{\sigma} = \phi_f \boldsymbol{\sigma}_f + \phi_s \boldsymbol{\sigma}_s.
\end{equation}
Since we are specifically concerned with incompressible materials in this paper, it is preferable to compute the pressure field as a separate independent variable, similar to mixed formulations in a Lagrangian context. For that purpose, one can consider a decomposition of the stress fields in the physical domains into their corresponding volumetric and deviatoric components. This step will allow us to have a single unified pressure field for the entire domain and two separate deviatoric stresses for the fluid and solid respectively as shown:
\begin{equation}
    \boldsymbol{\sigma} = -p \boldsymbol{I} + \phi_f \boldsymbol{\sigma}_f' + \phi_s \boldsymbol{\sigma}_s',
\end{equation}
where $p$ is the unified pressure field and $(')$ indicates deviatoric component of the stress tensor, given by $\boldsymbol{A}'=\boldsymbol{A} - \frac{1}{d} tr(\boldsymbol{A}) \boldsymbol{I}$ for a tensor $\boldsymbol{A}$, where $d$ is the dimension of the problem. The deviatoric components in the fluid and solid domains can be evaluated as follows:
\begin{equation}
    \boldsymbol{\sigma}_f' = \mu_f (\nabla \boldsymbol{v}_f + (\nabla \boldsymbol{v}_f)^T),
\end{equation}
and
\begin{equation}
    \boldsymbol{\sigma}_s' = \mu_s (\nabla \boldsymbol{v}_s + (\nabla \boldsymbol{v}_s)^T) + \mu_s^L (\boldsymbol{B} - \frac{1}{d} tr(\boldsymbol{B}) \boldsymbol{I}),
\end{equation}
where we have used the incompressible Neo-Hookean model for the hyperelastic solid. In the above equations, $\mu_f$ is the fluid viscosity, $\mu_s$ is the solid visocsity and $\mu_s^L$ is the shear modulus of the solid. Such a decomposition ensures that the definition of the Lagrange multiplier for enforcing incompressibility i.e. pressure is uniform across the domain. During the numerical implementation, we need to compute the Jacobian of the above stress term with respect to velocity to plug in the continuum equation. To achieve this, we utilize the transport equation of the left Cauchy-Green deformation tensor $\boldsymbol{B}$ as elaborated in \cite{mao2023interface}. We discuss more on Jacobian computations in \ref{appendixA}.
This completes the system of equations for the fluid-solid continuum. We focus on the implicit interface capturing method via phase-field functions in the next sub-section.

\subsection{Diffuse Interface Representation}\label{phase-field}
We utilize the phase-field model as a mathematical construct to evolve the interface implicitly during the FSI simulation. The phase-field function/ order parameter $\phi(\boldsymbol{x},t)$ varies steeply within the interface and has constant values outside of it. 
The Allen-Cahn equations for phase-field description can be derived from the process of free energy minimization, where the Ginzburg-Landau free energy functional is given by:
\begin{equation}
	\mathcal{E}: \mathcal{H}^1 (\Omega) \cap \mathcal{L}^4 (\Omega) \to \mathbb{R}_{\geq 0}, \mathcal{E}(\phi(\boldsymbol{x},t)) = \int_\Omega \left( F(\phi(\boldsymbol{x},t)) + \frac{\varepsilon^2}{2} |\nabla \phi(\boldsymbol{x},t) |^2 \right) d\Omega , 
\end{equation}
where $\Omega$ is the bounded physical domain, $H^1(\Omega)$ denotes the space of square-integrable real-valued functions with square-integrable derivatives on $\Omega$, $L^4(\Omega)$ denotes the function space in which the fourth power of the function is integrable, $\mathbb{R}_{\geq 0}$ represents the set of non-negative real numbers. The first term in the RHS is the bulk or mixing energy and depends on the local composition of the mixture. The functional form for the bulk energy $F(\phi)$ has been chosen as the double-well potential function $F(\phi) = \frac{1}{4} (\phi^2 - 1)^2$ in the present study. Thus, the value of the phase-field function within a pure solid and fluid domain is given by $1$ and $-1$ respectively. The second term is the interfacial or gradient energy and depends on the composition of the immediate environment. The ratio of these two effects controlled by $\varepsilon$ decides the thickness of the diffused interface region. At equilibrium, the interface thickness is the distance over which $\phi$ varies from $-0.9$ to $0.9$ which can be estimated as $2\sqrt{2}\tanh^{-1} (0.9) \varepsilon \approx 4\varepsilon$. 
The final convective form of the Allen-Cahn equation after addition of the Lagrange multiplier for mass conservation can be derived as follows:
\begin{equation}
	\frac{\partial \phi}{\partial t} + \boldsymbol{v} \cdot \nabla \phi = -\gamma \left ( F'(\phi) - \varepsilon^2 \Delta \phi - \beta (t) \sqrt{F(\phi)} \right ) \text{  on  } \Omega \times [0,T] ,
\end{equation}
where $\gamma$ is the mobility parameter (of the order $\mathcal{O}(10^{-3})$ for all the test problems in this work) and $\beta(t)$ is the time-dependent part of the Lagrange multiplier, given by $\beta(t) = \frac{\int_\Omega F'(\phi) d\Omega}{\int_\Omega \sqrt{F(\phi)} d\Omega}$. 

For the variational form of the Allen-Cahn equation, we define $\mathcal{S}^h$ to be the space of trial solutions, whose values satisfy the Dirichlet boundary conditions and $\mathcal{V}^h$ to be the space of test functions whose values vanish on the Dirichlet boundary. The stabilized form of the Allen-Cahn equation can be stated as: find $\phi_h^{n+\alpha} \in \mathcal{S}^h$ such that $\forall w_h \in \mathcal{V}^h$,
\begin{equation} \label{AC_var}
	\begin{aligned}
        \int_{\Omega}( w_{\mathrm{h}} \partial_{\mathrm{t}} \phi_{\mathrm{h}}^{\mathrm{n}+\alpha_m} + w_{\mathrm{h}} ( \boldsymbol{v}^{\mathrm{n}+\alpha} \cdot \nabla \phi^{\mathrm{n}+\alpha}_{\mathrm{h}}) &+ \gamma ( \nabla w_{\mathrm{h}} \cdot (  \varepsilon^2 \nabla \phi_{\mathrm{h}}^{\mathrm{n}+\alpha} ) +w_{\mathrm{h}} s \phi_{\mathrm{h}}^{\mathrm{n}+\alpha} - w_{\mathrm{h}} f ) ) \mathrm{d} \Omega   \\      
		&+\sum_{\mathrm{e}=1}^{\mathrm{n}_{\mathrm{el}}} \int_{\Omega_{e}}\left(\boldsymbol{v}^{\mathrm{n+\alpha}} \cdot \nabla w_{\mathrm{h}}\right) \tau_{\phi} \mathcal{R}(\phi_{\mathrm{h}}) \mathrm{d} \Omega_{e} = 0,			
	\end{aligned}
\end{equation}
where $s$ and $f$ are the reaction coefficient and source terms respectively and $\mathcal{R}\left(\phi_{\mathrm{h}}\right)$ is the element-wise residual of the Allen-Cahn equation as defined in \cite{joshi2020variational}. The first line of Eq. (\ref{AC_var}) contains the Galerkin terms, and the second line contains the linear stabilization terms with the stabilization parameter, $\tau_{\phi}$ \cite{shakib1991new, johnson2012numerical} given by
\begin{equation}
	\tau_{\phi} = \left[\left(\frac{2}{\Delta t}\right) + \boldsymbol{v} \cdot \boldsymbol{G}\boldsymbol{v} + 9\varepsilon^4 \boldsymbol{G}: \boldsymbol{G} + s^2 \right]^{-1/2} .
\end{equation}
The boundary condition and initial condition for the Allen-Cahn equation are as follows:
\begin{align}	
		\frac{\partial \phi}{\partial n} \biggr\rvert_{\Gamma} = \boldsymbol{n} \cdot \nabla \phi = 0 \ \ \ \  &\text{on   } \Gamma \times [0,T] ,\\
		\phi|_{t=0} = \phi_0 \ \ \ \  &\text{on   } \Omega .	
\end{align}
We define the initial phase field function $\phi_0$ based on the geometry of the solid bodies. This completes the set of the governing equations for the system. We now turn our attention to the central theme of the paper i.e. the contact modeling in our phase-field based FSI framework.

\section{Mono-field Interface Advancing Method for Collision Detection} \label{sec3}

\begin{figure}
    \centering
    \includegraphics[scale=0.7]{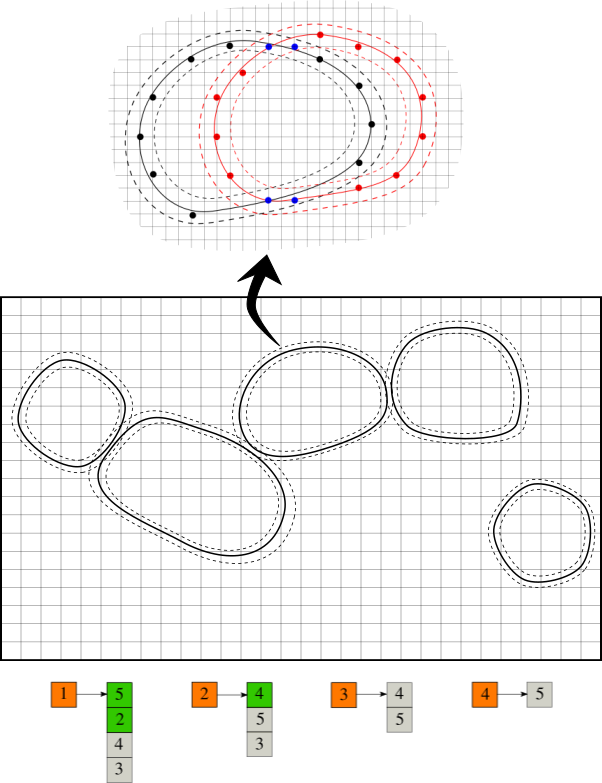}
    \put(-150,100){$\Omega_f$}
    \put(-230,160){$\Omega_s^1$}
    \put(-170,210){$\Omega_s^2$}
    \put(-40,135){$\Omega_s^3$}
    \put(-90,215){$\Omega_s^4$}
    \put(-305,205){$\Omega_s^5$}
    \put(-220,305){$\mathcal{B}_2^n$}
    \put(-120,320){$\mathcal{B}_2^{n+1}$}
    \caption{Illustration of mono-field interface advancing method for collision detection. The above schematic shows five different solid bodies ($N=5$) interacting and colliding in the surrounding fluid. The zoomed in image at the top shows the boundary update stage for $\Omega_s^2$ as it changes from $t^n$ to $t^{n+1}$. The lists at the bottom denote the sorted rank lists for each object, where the number $i$ stands for body $\Omega_s^i$. Orange-green pairs of objects are passed to the contact force subroutine as the distance between them ($g_{ij}$) is less than the specified threshold, whereas the orange-gray pairs are skipped while traversing the lists.}    
    \label{fig:mb_contact}
\end{figure}

In this section, we present our novel mono-field interface advancing (MFIA) method for simulating multibody contact scenarios while using a single order parameter. For solids with the same properties undergoing contact, e.g. ice floes on the water surface, this method can significantly reduce the complexity of the problem. It is important to distinguish between the different solid bodies present in the domain to identify the pairs of bodies that undergo contact at a particular instant. The traditional approach for such a problem would be to assign individual objects with their own order parameter field and strain/deformation field and solve these additional equations for the N solid bodies at every time step. However, the aforementioned approach can be computationally expensive with an increase in the number of bodies. 
Instead of reconstructing the individual interfaces at every time step, we now update the boundary node lists for each object instantiated at $t=0$.
This update process allows us to get away with one interface capturing field and one equation for the strain evolution for the N solid bodies in the physical domain. We then create rank lists for each body to detect possible collisions among any pair of bodies. Finally, we compute the contact forces among all the applicable pairs of solid bodies. We will discuss these three important stages of the proposed method in the sub-sections below. For the following discussions, let us consider that we have N objects that make up the solid domain $\Omega_s$, i.e. $\Omega_s = \cup_{i=1}^N \Omega_s^i$ and one fluid phase $\Omega_f$ for simplicity. 

\subsection{Update Process of Solid Boundaries}
We initialize the phase-field function of each solid body based on its location in the domain and isolate its boundary nodes $(\mathcal{B}_i^0)$ in the mesh at $t=0$ based on their $\phi$ values such that $-0.3 \le \phi(\mathcal{B}_i^0) \le 0.9$. 
When two bodies come close to each other, the asymmetric definition of boundary nodes allows to assign the mesh node to the correct body based on its distance.  
The first stage of the procedure involves the update process of the boundary node lists at each time step. At every time $t=t^n$ and for each node list $\mathcal{B}_i^n$, we check for nodes, $\mathcal{B}_{i,j}^n$ that no longer belong in the interface zone ($\phi(\mathcal{B}_{i,j}^n)<-0.3$ or $\phi(\mathcal{B}_{i,j}^n)>0.9$) and remove them from the list. Let us consider these set of nodes as dead nodes (black nodes in the top image of Fig. \ref{fig:mb_contact}), $\mathcal{D}_i$, where $i$ is the body number.

We then perform a domain scan for the phase indicator values of the nodes and collect the set of nodes $\Gamma$ that belong to the interface zone at the step ($-0.3 \le \phi(\Gamma) \le 0.9$). Using the above process, we create a new set $S$ such that $S = \Gamma - (\cup_i \mathcal{B}_i^n)$. 
The set $S$ contains all the new nodes that satisfy the interface zone criterion and need to be allocated to the correct boundary list, $\mathcal{B}_i$.
We achieve this by computing the distances of the particular node to the different bodies and finding the minimum among them. For the distance computations, we trim the boundary nodes to a tighter tolerance to create a set $\mathcal{I} \subset \mathcal{B}_i$ such that $|\phi(\mathcal{I}_{i,j})| \leq 0.1$ for all nodes indexed by $j$ in the set $\mathcal{I}_i$. This approximately isolates the iso-contours with $\phi=0$ for all the bodies. The tolerance can be adjusted based on the resolution for the particular simulation to ensure that we have a sufficient number of nodes to represent the contour with $\phi=0$. It is worth mentioning that the process does not involve an exact reconstruction of the sharp interface nonetheless it gives an approximate representation for the interface boundary. Thus, the distance between body $i$ and a node in the set $\mathcal{S}$ say $\mathcal{S}_k$ is given by $\mathrm{min}_j(\mathcal{S}_k - \mathcal{I}_{i,j})$. The process gives rise to the set of newly added nodes for each body, $\mathcal{N}_i$ (red nodes in the top image of Fig. \ref{fig:mb_contact}). Thus, the updated boundary lists can be represented as:
\begin{equation}\label{eq:bdry_upd}
    \mathcal{B}_i^{n+1} = \mathcal{B}_i^n - \mathcal{D}_i + \mathcal{N}_i.
\end{equation}
The blue nodes in the top image of Fig. \ref{fig:mb_contact} denote the nodes that are common to the boundary lists at $t^n$ and $t^{n+1}$. This completes the boundary update step of the procedure.

\subsection{Rank List Creation}
The next stage deals with creating the rank list for each solid body based on its distance from the other bodies. The distance calculations are carried out as mentioned in the previous sub-section. For every solid body $\Omega_s^i$, we compute its minimum gap from other bodies and arrange them in the ascending order of distances $g_{ij}$ ($g_{ij} = \mathrm{min}(\mathcal{I}_{i} - \mathcal{I}_j)$), as illustrated in Fig. \ref{fig:mb_contact}. The rank list for a higher-numbered body does not contain smaller-numbered bodies as those pairs are already accounted for in the earlier lists. Finally, we pass the pair of bodies that satisfy the minimum distance criterion to the contact force sub-routine (orange-green pairs in Fig. \ref{fig:mb_contact}), discussed in the next sub-section.

\subsection{Contact Force Computation}

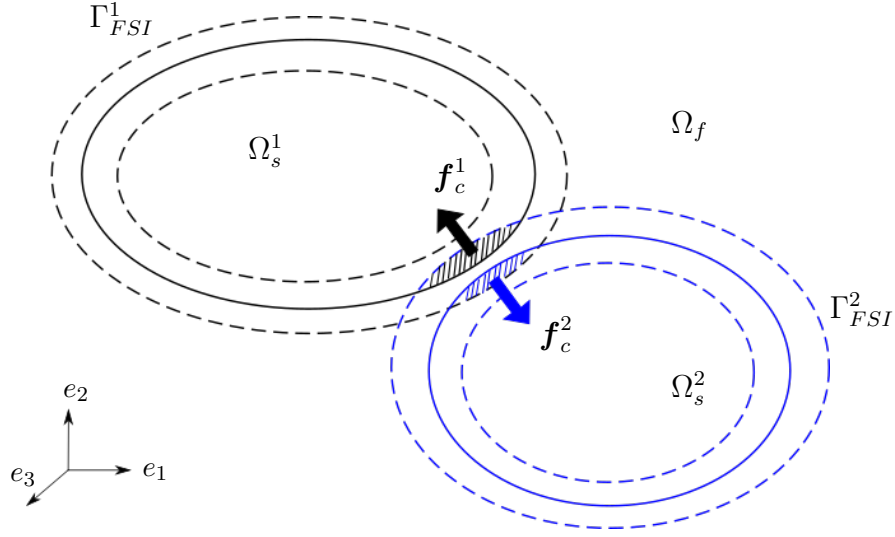
\begin{figure}
    \centering
    \includegraphics[scale=0.6]{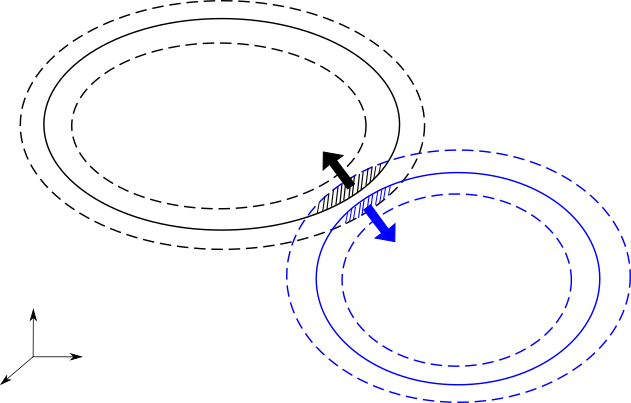}
    \put(-220,140){$\Omega_s^1$}
    \put(-60,50){$\Omega_s^2$}
    \put(-60,150){$\Omega_f$}
    \put(-280,190){$\Gamma_{FSI}^1$}
    \put(0,80){$\Gamma_{FSI}^2$}
    \put(-110,70){$\boldsymbol{f}_c^2$}
    \put(-150,130){$\boldsymbol{f}_c^1$}
    \put(-260,20){\small{$e_1$}}
    \put(-290,50){\small{$e_2$}}
    \put(-310,18){\small{$e_3$}}
    \caption{Schematic for the contact force between two colliding bodies. Two solid bodies $\Omega_s^1$ (black) and $\Omega_s^2$ (blue) with diffuse interfaces and immersed in a surrounding fluid ($\Omega_f$), approach contact as shown above. The hatched regions denote the area where the contact force is applied and the arrows represent the direction of the total contact force acting on the two bodies.}
    \label{fig:2body_force}
\end{figure}

As mentioned previously, we define a spring-type force as a consequence of two bodies coming into contact in the present model. Any further approach of the two solids beyond the transition zone is penalized by the force function. The treatment acts as a substitute for the reaction force that would result from a collision. Furthermore, this force is a function of the amount of overlap of the diffused zones (Fig. \ref{fig:2body_force}) and the physical and numerical parameters of the problem. The formulation is closest to the conventional penalty method for contact handling in terms of implementation. However, unlike the penalty method, the scope of inter-penetration or actual overlap between the solid bodies is much less in the current approach. 

Once a pair of bodies have been passed to this sub-routine, we loop over all the points in their respective boundary lists and compute the net exerted forces at the appropriate nodes. The force has a maximum at the mid-plane between the bodies and decays gradually.
The spring-type force function for our case is defined as:
\begin{align}
    \psi(x) = \left( 1-\frac{x}{2\varepsilon} \right),
\end{align}
where $4\varepsilon$ is the thickness of the diffuse interface. 
The vector-valued contact force can then be defined as:
\begin{align}\label{eq:f_contact}
    \boldsymbol{f}_c(d_{12}) = \kappa \mu_s^L \psi(d_{12}) \boldsymbol{n},
\end{align}
where $\kappa$ is a constant used to control the magnitude of the force and $\boldsymbol{n}$ is the common unit normal defined as $\boldsymbol{n} = \pm \frac{\nabla d_{12}}{|\nabla d_{12}|}$. 
In the above equation, $d_{12}$ is the relative level set of a point from the two bodies, which is defined as:
\begin{align}
    d_{12} = \frac{d_1-d_2}{2},
\end{align}
where $d_1$ and $d_2$ are the minimum distances (level set) of the point from the approximate zero iso-contours ($\mathcal{I}_i$) of the two bodies as defined in the previous sub-section. For example, for a point in the boundary of body 1 ($P \coloneqq \mathcal{B}_{1,k}$), these distances are given by $d_1 = \mathrm{min}_j(P - \mathcal{I}_{1,j})$ and $d_2 = \mathrm{min}_j(P - \mathcal{I}_{2,j})$. We note that $d_{12}=0$ denotes a point equidistant from the two bodies. 

The computed contact force is then incorporated in the body force term in the continuum equation (Eq. \ref{eq:UC}). In problems involving gravity, the total body force is given as:
\begin{equation}
    \boldsymbol{b}(\phi) = \rho(\phi) \boldsymbol{g} + \boldsymbol{f}_c.
\end{equation}
We also account for the collision of an object with the domain boundaries (rigid walls), where the distance is computed between the $\phi=0$ boundary of the body to the rigid wall and the unit normal of the fixed wall is utilized for the force direction. 
The computed forces of each pair are then included in the unified momentum balance equation. For example, with reference to Fig. \ref{fig:mb_contact}, forces between pairs 1-5 and 1-2 are computed separately and included in the one-field momentum equation.

\subsection{MFIA Algorithm}
We summarize the discussions from the previous sub-sections in Algorithm \ref{algorithm_1}.

\begin{algorithm}
	\caption{Mono-field interface advancing (MFIA) method for collision detection}
	\label{algorithm_1}
	\begin{algorithmic}[1]
	\STATE	Initialize phase-field functions and boundary lists $\mathcal{B}_i^0$ for every solid body \\
	\STATE Loop over time steps $\mathrm{n}=1,2,\cdots$ \\	
		\STATE \quad Loop over each boundary list $\mathcal{B}_i^n$  \\
	  \STATE \qquad Remove nodes ($\mathcal{D}_i$) that do not lie in the interface zone \\
		\STATE \quad Loop over all nodes and find the ones that are in the interface zone at $t=t^{n+1}$ \\
		\STATE \quad Extract the set of nodes that have recently entered the interface zone \\
		\STATE \quad Assign these nodes ($\mathcal{N}_i$) to the correct boundary list $\mathcal{B}_i$  \\
		\STATE \quad Loop over the solid bodies  \\
		\STATE \qquad Create the rank list based on the proximities from other bodies  \\
		\STATE \qquad Obtain the pairs of potentially colliding bodies based on the minimum gap criterion \\
		\STATE \qquad Pass these pairs to the contact force sub-routine (Eq. \ref{eq:f_contact}) \\		 	
	\end{algorithmic}
\end{algorithm}

The novel MFIA approach allows us to simulate multiple bodies of identical physical parameters immersed in a surrounding fluid. It eliminates the need for solving N number of level set/ phase field equations and solid strain equations for each of N bodies. This also prevents the creation of new lists at each time step by tweaking the existing lists by a small amount. The above approach can be considered similar to the fast marching method used in level set problems \cite{sethian1996fast}. The fast marching method involves solving the convection equation for the level set in a narrow band around the existing front in an upwind fashion (or normal to the front) and maintaining the values for the far-away points. However, in the current approach, we solve the unified phase-field equation in the full domain to maintain the thermodynamic consistency and only carry out the procedure described above to distinguish between the different bodies. The assumptions involved are that the mesh size and time step size are sufficiently small to avoid excessive displacements, which are reasonable for any contact simulation. 

\begin{remark}
    Another related approach is the interface reconstruction technique in the cut-cell method or in X-FEM which is carried out to maintain the sharp interface nature of the method \cite{parvizian2007finite, wall2008fluid}. We would like to highlight, however, that in the present approach, we are not concerned with the location of the exact sharp interface. We are only interested in separating the updated diffuse regions of each body after solving the Allen-Cahn equation. Thus, the above procedure does not pertain to any form of interface or geometry-preserving ideas \cite{mao2023interface} but can be used in conjunction with one.
\end{remark}

\begin{remark}
    The proposed method avoids the use of any material/ Lagrangian points as in the particle level set \cite{enright2002hybrid} or additional indicator functions as in the front tracking methods \cite{unverdi1992front} and relies entirely on the Eulerian mesh nodes. 
\end{remark} 

\noindent This completes the description of our Eulerian phase-field based contact formulation. We will now present some implementation details of the variational finite element solver before turning our attention to verification and applications.

\section{Numerical Procedure} \label{sec5}
In this section, we present the details of the numerical implementation of the proposed framework in our in-house solver \cite{mao2023interface, rath2023interface}.

\subsection{Temporal Discretization}
The temporal discretization is carried out via the generalized-alpha time integration method \cite{chung1993time, jansen2000generalized}. It provides a single user-controlled parameter called the spectral radius $\rho_{\infty}$ to dampen the undesirable high-frequency oscillations in the solution. The generalized-$\alpha$ method for any generic variable $\varphi$ can be given as
\begin{align}
	\varphi^{n+1}&=\varphi^{n}+\Delta t \partial_t \varphi^n +\Delta t\varsigma (\partial_t \varphi^{n+1}-\partial_t \varphi^n),\\
	\partial_t \varphi^{n+\alpha_m} &= \partial_t \varphi^{n}+\alpha_m (\partial_t \varphi^{n+1}-\partial_t \varphi^n),\\
	\varphi^{n+\alpha}&=\varphi^n+\alpha(\varphi^{n+1}-\varphi^{n}),
\end{align}
where $\Delta t$ is the time step size, $\alpha_m$, $\alpha$ and $\varsigma$ are the generalized-$\alpha$ parameters defined as:
\begin{align}
	\alpha=\frac{1}{1+\rho_{\infty}},\ \alpha_m=\frac{1}{2}\left(\frac{3-\rho_{\infty}}{1+\rho_{\infty}}\right),\ \varsigma=\frac{1}{2}+\alpha_m-\alpha.
\end{align} 
The above method works in a predictor-multicorrector type technique between time steps $t^{n+1}$ and $t^{n+\alpha}$. For this study, we set $\rho_{\infty}=1$ in the simulations, which essentially recovers to the trapezoidal time integration. 

\subsection{Implementation Details}
We first initialize the velocity, pressure, order parameter and deformation tensor in the domain and apply the boundary conditions depending on the problem at hand. We initialize the boundary lists $\mathcal{B}_i^0$ for each body $\Omega_s^i$ at $t=0$. The boundary nodes on each wall of the domain are stored in separate arrays in addition to the above lists and the wall normals are computed and saved. 
The solver then enters the time loop where the non-linear set of equations is solved in a partitioned manner \cite{rath2023interface}. Inside the time loop, we first check for potentially colliding bodies and impose the appropriate contact forces before entering into the non-linear Newton iterations. The Jacobians are calculated, and the matrices are assembled for each of the equations separately and then solved sequentially. 
The linearized system for the unified continuum equations can be formulated as:
\begin{align} \label{LS_UCeq}
	\begin{bmatrix}
		\boldsymbol{K}_\Omega &  & \boldsymbol{G}_\Omega \\
		& \\
		-\boldsymbol{G}^T_\Omega &  &\boldsymbol{C}_\Omega
	\end{bmatrix} 
	\begin{Bmatrix}
		\Delta \boldsymbol{v}^\mathrm{n+\alpha} \\
		\\
		\Delta p^\mathrm{n+1}
	\end{Bmatrix}
	= \begin{Bmatrix} 
		-\widetilde{\boldsymbol{\mathcal{R}}}_\mathrm{m}(\boldsymbol{v},p) \\
		\\
		-\widetilde{\mathcal{R}}_\mathrm{c}(\boldsymbol{v})
	\end{Bmatrix} ,
\end{align}
where $\boldsymbol{K}_\Omega$ is the stiffness matrix of the unified momentum equation consisting of inertia, convection, diffusion and stabilization terms, $\boldsymbol{G}_\Omega$ is the discrete gradient operator, $\boldsymbol{G}^T_\Omega$ is the divergence operator and $\boldsymbol{C}_\Omega$ is the pressure-pressure stabilization term. Here, $\Delta \boldsymbol{v}$ and $\Delta p$ are the increments in the velocity and pressure, respectively, and $\widetilde{\boldsymbol{\mathcal{R}}}_\mathrm{m}(\boldsymbol{v},p)$ and $\widetilde{\mathcal{R}}_\mathrm{c}(\boldsymbol{v})$ represent the weighted residuals of the stabilized momentum and continuity equations, respectively. Let the updated quantities at $t^\mathrm{n+1}_\mathrm{(k)}$ be represented as $\boldsymbol{X}^\mathrm{n+1}_\mathrm{(k)}$, $\mathrm{k}$ being the nonlinear iteration index, and the increments in these quantities be represented as $\Delta\boldsymbol{X}$. The error in solving the unified continuum equations can be defined as
\begin{align} \label{e_UC}
	e_{UC} = \frac{||\Delta \boldsymbol{X}||}{||\boldsymbol{X}^\mathrm{n+1}_{\mathrm{(k)}}||}.
\end{align}  
The updated velocities from the solution of the unified continuum equations are used for evolving the solid strains via the $\boldsymbol{B}$ transport equation and updating the order parameter values through the Allen-Cahn equation.
The linearized form for the transport of the left Cauchy-Green tensor is given by:
\begin{align} \label{LS_CGT}
	\begin{bmatrix}
		\boldsymbol{K}_{CGT}
	\end{bmatrix} 
	\begin{Bmatrix}
		\Delta \boldsymbol{B}^\mathrm{n+\alpha}
	\end{Bmatrix}
	= \begin{Bmatrix} 
		-\widetilde{\mathcal{R}}(\boldsymbol{B})
	\end{Bmatrix} ,
\end{align}
where $\boldsymbol{K}_{CGT}$ is the stiffness matrix and $\widetilde{\mathcal{R}}(\boldsymbol{B})$ represents the weighted residual for the left Cauchy Green tensor equation. The numerical error in solving the above equation can be written as
\begin{align} \label{e_CGT}
	e_{CGT} = \frac{||\Delta \boldsymbol{B}^\mathrm{n+\alpha}||}{||\boldsymbol{B}^\mathrm{n+1}_{\mathrm{(k)}}||}.
\end{align} 
Solid stresses are computed from the new deformation tensor $(\boldsymbol{B})$ values.
Similarly, the linearized form for the Allen-Cahn equation can be expressed as
\begin{align} \label{LS_AC}
	\begin{bmatrix}
		\boldsymbol{K}_{AC}
	\end{bmatrix} 
	\begin{Bmatrix}
		\Delta \phi^\mathrm{n+\alpha}
	\end{Bmatrix}
	= \begin{Bmatrix} 
		-\widetilde{\mathcal{R}}(\phi)
	\end{Bmatrix} ,
\end{align}
where $\boldsymbol{K}_{AC}$ consists of the inertia, convection, diffusion, reaction and stabilization terms and $\widetilde{\mathcal{R}}(\phi)$ represents the weighted residual for the stabilized conservative Allen-Cahn equation.  The numerical error in solving the Allen-Cahn equation can be written as
\begin{align} \label{e_AC}
	e_{AC} = \frac{||\Delta \phi^\mathrm{n+\alpha}||}{||\phi^\mathrm{n+1}_{\mathrm{(k)}}||}.
\end{align} 
The updated order parameter values are utilized further in the interpolation of density $\rho(\phi)$, viscosity $\mu(\phi)$ and stress $\boldsymbol{\sigma}(\phi)$. The new phase-field functions also help update the solid boundaries and estimate the gap between the solid bodies during the broad-phase collision detection step. The pairwise contact forces obtained for the various elastic-elastic and elastic-rigid collisions are computed based on the amount of overlap of the diffuse interfaces. We evaluate the total contact force acting on a given body as a vector sum of all the collision forces acting on it. These additional forces that contribute to the body force along with the other updated physical quantities are fed to the unified continuum equation for the next iteration.
P1 triangular elements and Q1 quadrilateral elements have been used for the various test cases presented in this paper. The convergence criterion for the non-linear loop is based on the norm of the solution ratios mentioned above ($e_{UC}$, $e_{CGT}$ and $e_{AC}$). We exit the non-linear loop when these error norms go below $10^{-4}$ or when the maximum number of non-linear iterations (10 in our case) is reached.

\section{Verification and Convergence} \label{sec6}
We provide a systematic verification of the above contact force implementation for a dry contact scenario and an FSI contact scenario in this section. We utilize the analytical Hertzian contact solutions for dry contact or solid-only contact verification. We plot the traction profile over the contact patch and force-indentation curves for this example and compare them with available analytical solutions. We also present the convergence of the implemented contact routine with respect to the mesh resolution for a fixed interface resolution ($\frac{\varepsilon}{h}$). 
For the FSI contact verification, we solve the problem of two-body collision in a Taylor-Green vortex field and present a comparison of the centroid trajectories with existing literature.  

\subsection{Dry Contact Verification}
The analytical solutions available for smooth Hertzian contact between two linear elastic bodies are utilized for the assessment of our phase-field model in dry contact scenarios.
The Hertzian contact between two cylinders results in an elliptic stress distribution of width $2a$ across the contact zone. Let the cylinders have radii $R_1$ and $R_2$, Young's moduli $E_1$ and $E_2$ and Poisson's ratios $\nu_1$ and $\nu_2$ respectively. The half-width $a$ for this scenario is given by:
\begin{equation}
    a = \left( \frac{4FR_{eq}}{\pi E_{eq}} \right)^{\frac{1}{2}},
\end{equation}
where $F$ is the total contact force acting on each body, $R_{eq}$ is the equivalent radius of curvature, $R_{eq}=\frac{R_1 R_2}{R_1 + R_2}$ and $E_{eq}$ is the equivalent Young's modulus, $\frac{1}{E_{eq}} = \frac{(1-\nu_1^2)}{E_1} + \frac{(1-\nu_2^2)}{E_2}$. 
The maximum contact pressure is given by,
\begin{equation}
    p_0 = \frac{2F}{\pi a}.
\end{equation}
The elliptic traction profile is given by,
\begin{equation}
    \frac{p(r)}{p_0} = \left( 1 - (r/a)^2 \right)^{\frac{1}{2}} \text{       in      } 0\le r \le a.
\end{equation}

\noindent The above relations apply to the contact of a cylinder with a plane surface as well, where $R=\infty$ for the plane surface.

\subsubsection{Deformable Cylinder on Elastic Plane Surface}

\begin{figure}
    \centering
    \includegraphics[scale=0.6]{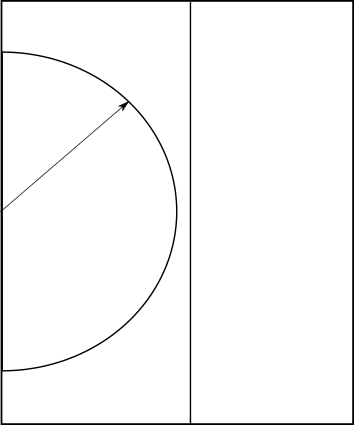}
    \put(-140,70){$\Omega_s^1$}
    \put(-50,70){$\Omega_s^2$}
    \put(-150,133){$R$}
    \put(7,90){$v_x = -u_0$}
    \caption{Schematic of a simple Hertzian contact problem with a deformable cylinder contacting an elastic plane. The plane wall on the right approaches the fixed cylinder on the left with a constant velocity boundary condition at the right boundary.}
    \label{fig:solid_contact}
\end{figure}

\begin{figure}[htbp] 
    \begin{subfigure}[h]{0.4\textwidth}
        \centering
        \includegraphics[width=1\textwidth]{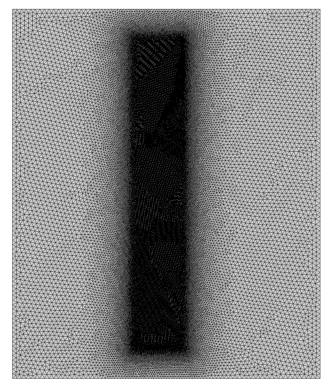}
        \caption{}
        \label{hertz_mesh}
    \end{subfigure}
    \begin{subfigure}[h]{0.45\textwidth}
        \centering
        \includegraphics[width=1\textwidth]{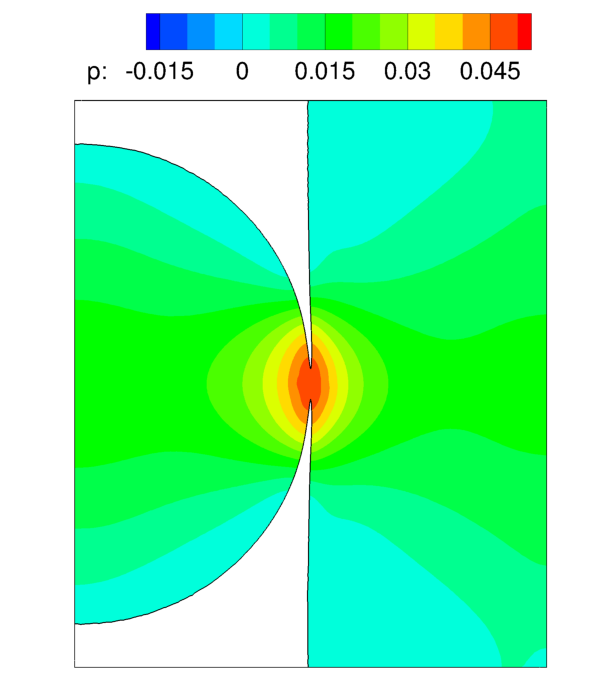}
        \caption{}
        \label{hertz_p}
    \end{subfigure}
    \begin{subfigure}[h]{0.51\textwidth}
        \centering
        \includegraphics[width=1\textwidth]{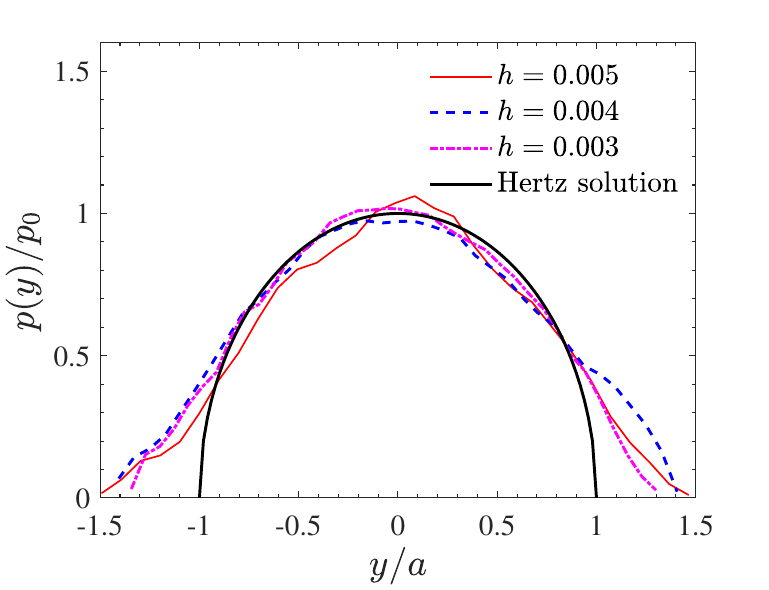}
        \caption{}
        \label{hertz_compare}
    \end{subfigure}
    \begin{subfigure}[h]{0.48\textwidth}
        \centering
        \includegraphics[width=1\textwidth]{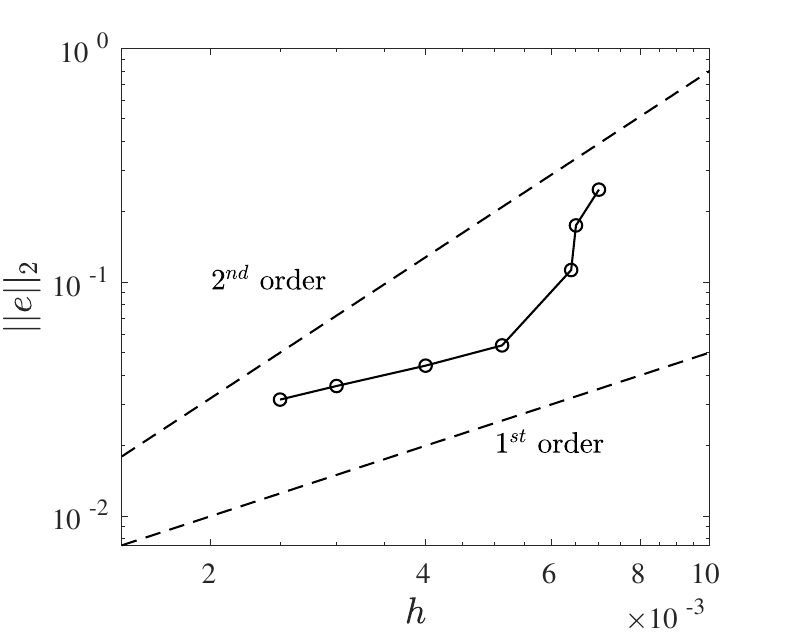}
        \caption{}
        \label{hertz_conv}
    \end{subfigure}
    \centering
    \caption{Hertz contact problem on an elastic plane: (a) mesh for the contact problem between a deformable cylinder on an elastic plane. The mesh is refined within the expected contact region. (b) Pressure contour for the contact problem, (c) normalized traction profiles for different mesh resolutions and their comparison with the analytical Hertz solution and (d) convergence of the relative L2 error in the traction profile with respect to $h$.  The numerical traction profiles plotted in (c) have been smoothed out by performing a moving average of the obtained data and the errors in (d) have been plotted for $\frac{p(y)}{p_0}>0.3$ to compare only the linear regime of non-conforming contact. The interface resolution is similar for all the cases, $\frac{\varepsilon}{h} \approx 10$.}
    \label{fig:hertz_compare}
\end{figure}

In this test case, we consider a deformable cylinder with $R=1$ that is compressed by an elastic plane (Fig. \ref{fig:solid_contact}). We model the second phase as air to emulate dry solid-solid contact. We use the incompressible Neo-Hookean model to represent both the solids and assign a Young's modulus of 1. Other physical properties are: $\rho_f=10^{-3}$, $\rho_s=1$, $\mu_f=10^{-4}$. All the physical parameters are non-dimensional in nature. We carry out the simulations for different mesh resolutions with $\kappa=30$ and a time step size of $0.02$. The center of the cylinder is located at (0,0). The initial condition for the order parameter is defined as:
\begin{equation}
    \phi(x,y,0) = 1 + \mathrm{tanh}\left(\frac{ R - \sqrt{(x-0)^2 + (y-0)^2}}{\sqrt{2}\varepsilon} \right) + \mathrm{tanh}\left( \frac{x-x_0}{\sqrt{2}\varepsilon}  \right).
\end{equation}
We initialize the elastic plane in the region $x_0 \le x \le 2$ so that it slowly approaches the cylinder and presses on it. The value of $x_0$ is slightly modified for different mesh resolutions ($x_0 \in [1.1,1.24]$) to ensure that the two bodies are sufficiently apart and do not penetrate each other at the start of the simulation. The traction profiles are extracted when the total contact force ($F$) on the cylinder reaches a prescribed value ($\approx 0.005$ in our case). 
The cylinder is fixed on the left boundary. A constant velocity of $u_0=R/100$ is applied at the right boundary to emulate the slow approach of the plane toward the cylinder. During post-processing, we compute the traction profile by integrating the contact body force $\boldsymbol{f}_c$ normal to the interface. The total contact force is given by the volume integral, $F = \int_{\Omega_{cyl}} \boldsymbol{f}_c \cdot \boldsymbol{n} \mathrm{d}\Omega$ for the cylinder. Numerically, the traction computation is done as follows. We first define a new variable that stores the quantity $f = \boldsymbol{f}_c \cdot \boldsymbol{n}$. We define a local rectangular grid around the contact patch to approximate the diffused contact region. We interpolate the values of the newly computed quantity over the grid and carry out quadrature integration over $x$-points for each $y$ location as $p(y) = \int_{x_1}^{x_2} f(x,y) dx$. This gives us the traction profile $p(y)$ along the contact patch.

Figure \ref{hertz_compare} shows the comparison of the obtained traction profiles with the analytical Hertz profile. In general, the numerical traction profiles follow the elliptic trend of the analytical solution. However, some deviations can be observed in the solutions which can be minimized by using better reconstruction approaches while computing the contact pressure. Some of the other possible causes for the deviation might be (i) the use of a diffused interface profile, which results in a volumetric contact force (area force in 2D) instead of a surface force (line force in 2D); (ii) an incompressible hyperelastic material law compared to a compressible linear elastic material in Hertz theory; and (iii) the finite size of the computational domain as compared to the semi-infinite assumption of Hertz. The difference in the material laws also gives us a much wider contact patch compared to the Hertzian solution. This is more similar to the case of conforming contact. We have plotted the convergence of the relative $L_2$ error in the traction profile with different mesh resolutions in the contact zone $h$ in Fig. \ref{hertz_conv}. For these error computations, we utilize the force profile within $\frac{p(y)}{p_0}>0.3$ to stay in the linear regime of non-conforming contact. We define the relative $L_2$ error in the obtained traction profile as:
\begin{equation}
    ||e||_2 = \frac{||p(y) - p_{Hertz}||_2}{||p_{Hertz}||_2},
\end{equation}
where $p_{Hertz}$ is the traction profile obtained via Hertz theory using the numerically computed total contact force $F$. We can observe that as we reduce the mesh size, the error in the traction profile decreases monotonically. The convergence rate is found to be between the first and second order ($\approx 1.8$).

\subsubsection{Force-indentation Curve}
We plot the force as a function of the indentation curve for the contact problem described in the previous section to verify the validity of the model for longer simulations  \cite{abuhattum2022explicit}. We utilize the analytical solutions from Hertz ($F_{Hertz}$) and Ding et al. \cite{ding2017determination} ($F_{Ding}$) to compare our numerical results in the linear and non-linear regimes of the problem:
\begin{equation}
    F_{Hertz} = \frac{4}{3} \frac{E_{eq}}{(1-\nu^2)} \sqrt{R_{eq}\delta^3},
\end{equation}
\begin{equation}
    F_{Ding} = F_{Hertz} \left(1-0.15 \frac{\delta}{R_{eq}} \right),
\end{equation}
where $\delta$ is the indentation depth and other symbols have their usual meaning as defined previously. 
\begin{figure}
    \centering
    \includegraphics[scale=0.8]{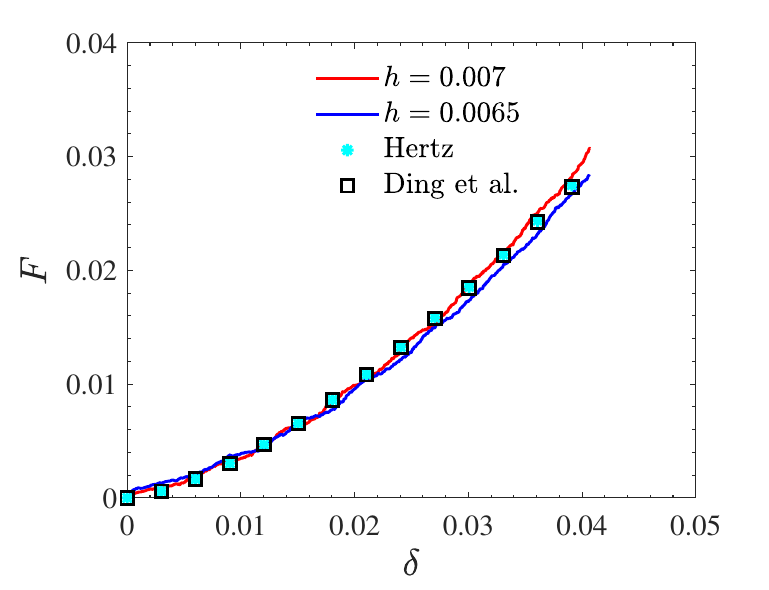}
    \caption{Comparison of the force vs indentation curves for contact between a deformable cylinder and an elastic plane. Results from two mesh resolutions are compared with the analytical solutions of Hertz theory and Ding et al. \cite{ding2017determination}.}
    \label{fig:f_delta}
\end{figure}
We keep the same material parameters and the interface resolution ($\frac{\varepsilon}{h}\approx10$) as before. We take the value of $\kappa$ as 1.87 with a time step size of $0.01$ and plot the results for two different mesh resolutions (Fig. \ref{fig:f_delta}). We obtain the total force ($F$) numerically by performing a volume integral over the cylinder as described in the previous sub-section. The force indentation curves are plotted once the bodies are in contact and the contact forces have been applied. We can observe that our numerical results match well with the analytical solutions. It is worth noting that it is not feasible to continue the simulations beyond a certain depth of indentation due to the incompressibility assumption and fixed boundary conditions for the cylinder on the left boundary. Thus, the Hertz and Ding solutions are nearly identical for the simulation duration.

\subsection{Elastic-Elastic Contact in Fluid Environment}\label{TG_2body}
To ensure the accuracy of the implemented model for generalized FSI contact cases, we present a simple verification of two hyperelastic bodies colliding in the Taylor-Green vortex field. We compare the trajectories of the centroids of the two bodies with other works in the literature and present discussions on the energy transfer mechanisms and volume conservation. Finally, we illustrate the cost-effectiveness of the proposed MFIA approach over a multi phase-field (MPF) approach for collision detection and contact handling using this test case.

\subsubsection{Two Body Collision in Taylor-Green Vortex Field}

\begin{figure}[htbp]
	\centering
	\begin{subfigure}{0.24\textwidth}
		\centering		
		\includegraphics[width=\textwidth]{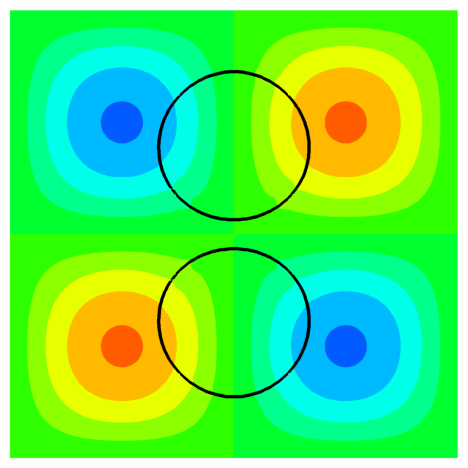}            
		\caption{}
	\end{subfigure}
	\hfill
	\begin{subfigure}{0.24\textwidth}
		\centering		
		\includegraphics[width=\textwidth]{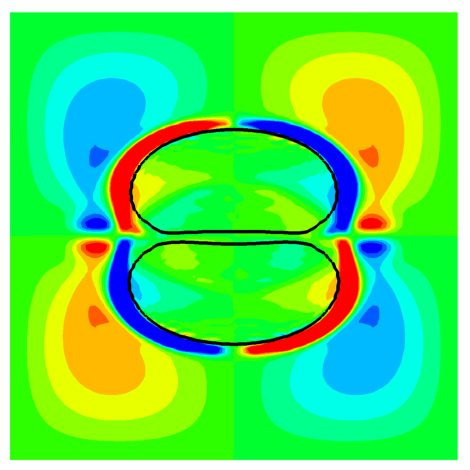}
		\caption{}
	\end{subfigure}
        \begin{subfigure}{0.24\textwidth}
		\centering		
		\includegraphics[width=\textwidth]{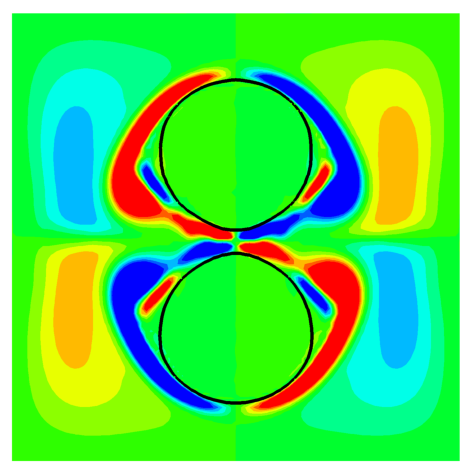}
		\caption{}
	\end{subfigure}
	\hfill
	\begin{subfigure}{0.24\textwidth}
		\centering		
		\includegraphics[width=\textwidth]{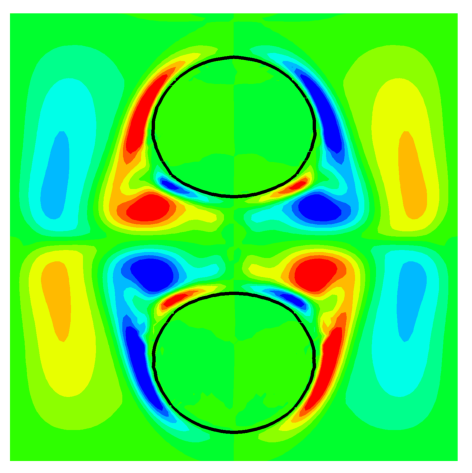}
		\caption{}
	\end{subfigure}

    \begin{subfigure}{0.35\textwidth}
		\centering		
		\includegraphics[width=\textwidth]{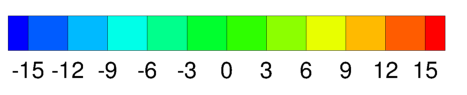}
        \put(-185,6){$\omega_z:$}
		\label{legend_TG_2body}
	\end{subfigure}   
	\caption{Two elastic body collision problem in a Taylor-Green voretx field. Snapshots of the vorticity contours for the two body collision problem at $t=$ (a) $0$, (b) $0.3$, (c) $0.6$ and (d) $0.9$. The black solid lines denote the iso-contours with $\phi=0$  i.e. the boundaries of the two bodies.}
	\label{fig:TG_2body_collision}
\end{figure}

\begin{figure}[htbp]
    \centering
    \begin{subfigure}{0.5\textwidth}
        \centering
        \includegraphics[width=\textwidth]{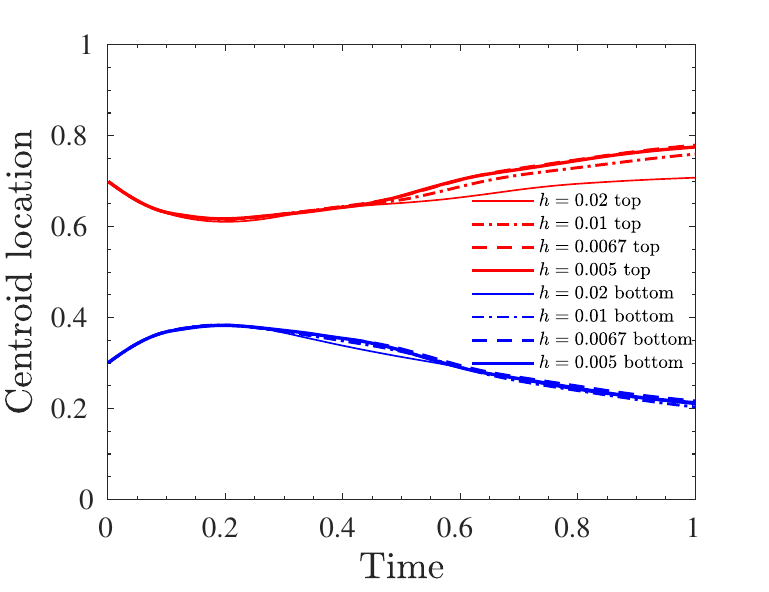}
        \caption{}
        \label{TG_y_mesh}
    \end{subfigure}
    \hfill
    \begin{subfigure}{0.48\textwidth}
        \centering
        \includegraphics[width=\textwidth]{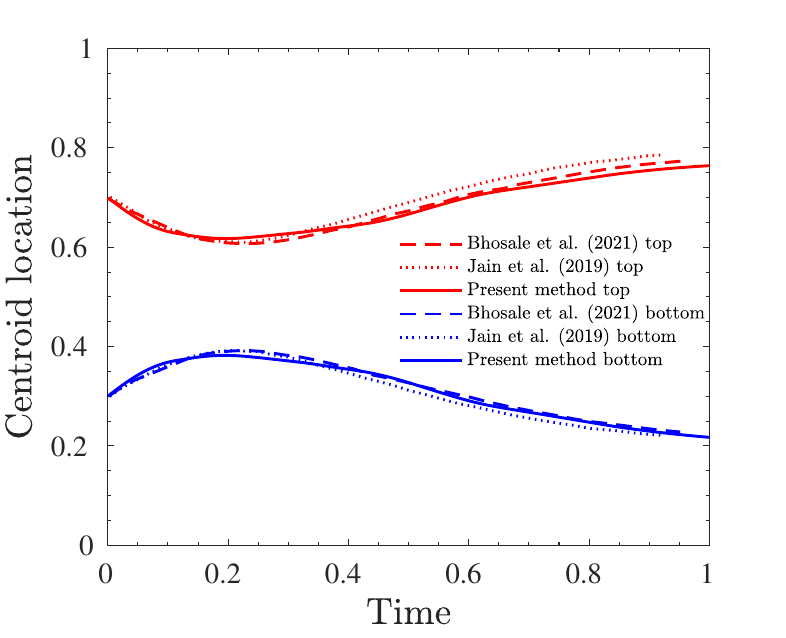}
        \caption{}
        \label{TG_y_paper}
    \end{subfigure}
    \caption{Two elastic body collision problem in a Taylor-Green voretx field. Time evolution of the centroids of the top and bottom bodies: (a) solution convergence with different mesh resolutions, $h$ and (b) comparison with previous works by Bhosale et al. (2021) \cite{bhosale2021remeshed} and Jain et al. (2019) \cite{jain2019conservative} for $h=0.005$ case.}
    \label{fig:TG_y_compare}
\end{figure}

To simulate this problem, we initialize two incompressible hyperelastic bodies in a Taylor-Green vortex field and allow them to collide and separate eventually. This problem has previously been performed in \cite{jain2019conservative, bhosale2021remeshed} to simulate soft solid contact. The non-dimensional geometric and physical parameters for the problem are as follows. We define a unit square over $[0, 1] \times [0, 1]$ as the computational domain for this simulation. Two incompressible solids that are initially circular with radii $R_1 = R_2 = 1/6$ are placed at $(0.5,0.7)$ and $(0.5,0.3)$. The phase-field function for the two-body system is initialized as follows:
\begin{equation}
\begin{aligned}
    \phi(x,y,0) = 1 &+ \mathrm{tanh}\left( \frac{R_1 - \sqrt{(x-0.5)^2 + (y-0.7)^2}}{\sqrt{2}\varepsilon} \right) \\
    &+ \mathrm{tanh}\left( \frac{R_2 - \sqrt{(x-0.5)^2 + (y-0.3)^2}}{\sqrt{2}\varepsilon} \right).
\end{aligned}
\end{equation}
The initial Taylor-Green vortex field given by the streamfunction $\psi = \psi_0 \mathrm{sin}(k_xx) \mathrm{sin}(k_yy)$ is imposed using the following velocity distribution in the domain:
\begin{align}
    v_x(x,y) &= \psi_0 k_y \mathrm{sin}(k_x x)\mathrm{cos}(k_y y) \\
    v_y(x,y) &= -\psi_0 k_x \mathrm{cos}(k_x x)\mathrm{sin}(k_y y),
\end{align}
where $\psi_0=\frac{1}{2\pi}$, $k_x=2\pi$ and $k_y=2\pi$ for the present case. Other physical parameters have been chosen the same as those in previous works (\cite{bhosale2021remeshed}) with fluid viscosity $\mu_f=\frac{0.01}{2\pi}$, shear modulus $\mu_s^L=2$ and density $\rho_f=\rho_s=1$. The interface resolution is considered to be $\frac{\varepsilon}{h}=1$ and the time step size is 0.001 for all simulations of this test case. The force constant $\kappa$ is assumed to be $1.5$. We impose the periodic boundary conditions on all sides for the momentum equation and homogeneous Neumann conditions for the phase-field equation. Figure \ref{fig:TG_2body_collision} shows the temporal evolution of the centroids of the two solids as they approach collision and separate eventually.

\subsubsection{Assessment of Volume Conservation and Energy Transfer}
\begin{figure}[htbp]
    \centering
    \begin{subfigure}{0.49\textwidth}
        \centering
        \includegraphics[width=\textwidth]{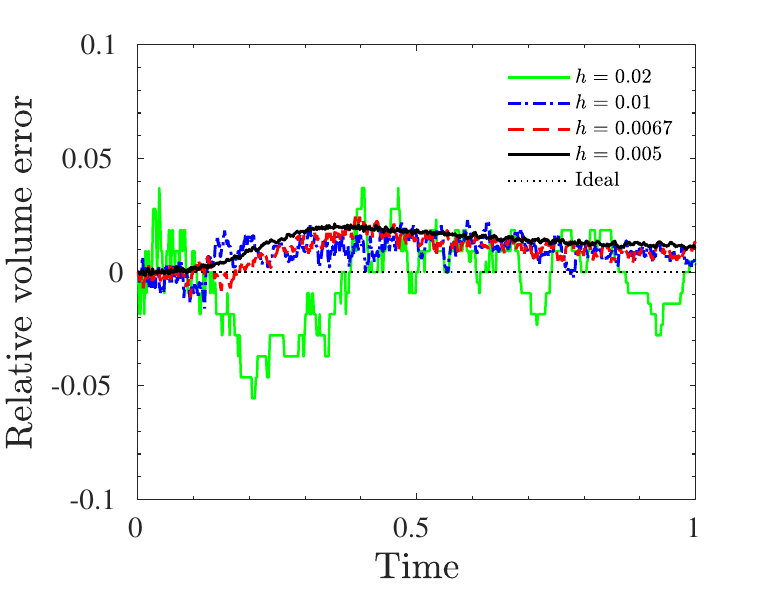}
        \caption{}
        \label{TG_vol_mesh}
    \end{subfigure}
    \hfill
    \begin{subfigure}{0.49\textwidth}
        \centering
        \includegraphics[width=\textwidth]{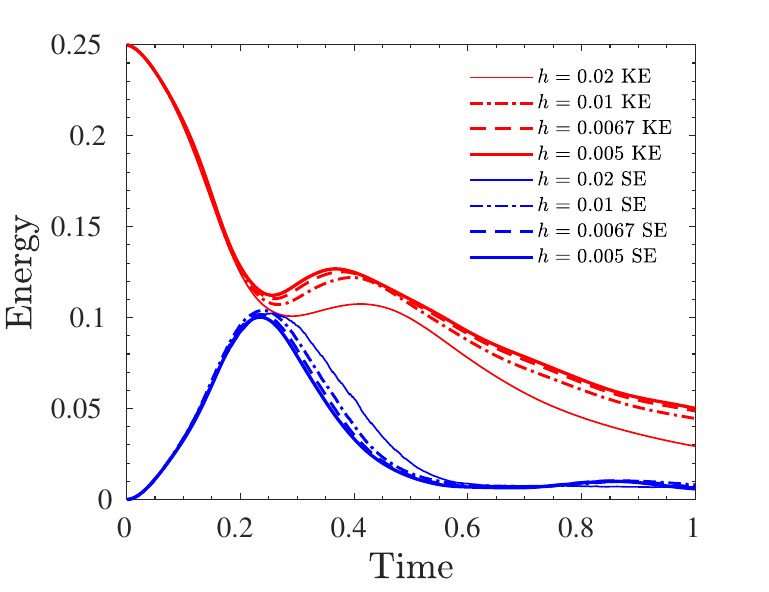}
        \caption{}
        \label{TG_energy_mesh}
    \end{subfigure}
    \caption{Comparison of temporal evolution of (a) relative volume error for the top body and (b) kinetic and strain energies of the system for different mesh resolutions. The dashed black line in the left figure shows the ideal case of constant volume.}
    \label{fig:TG_mesh_compare}
\end{figure}

We measure the error in volume conservation of the solids during the collision and separation processes for the present case. For an incompressible solid, this quantity gives us an estimate of the deviation of $J=\mathrm{det}(\boldsymbol{F})$ from unity. We define the relative volume error for the top body using the following definition:
\begin{equation}
    e_{vol} = \frac{\int_{\Omega_{s,1}^t}\mathrm{d}\Omega - \int_{\Omega_{s,1}^0}\mathrm{d}\Omega}{\int_{\Omega_{s,1}^0}\mathrm{d}\Omega}.
\end{equation}
We plot this quantity for different mesh resolutions in Fig. \ref{TG_vol_mesh}. It can be concluded that the magnitude of the relative volume error is within $5\%$ for all the cases and within $2\%$ for the finer resolution cases ($h\le0.01$). The deviation from the ideal scenario of constant volume can be attributed to a lower interface resolution ($\frac{\varepsilon}{h} = 1$) of the diffused interface body. This issue can be alleviated by the application of adaptive mesh refinement techniques as shown in \cite{rath2023interface}.

We also track the evolution of energy for this test case to illustrate the transfer of kinetic energy to solid strain energy and vice versa (Fig. \ref{TG_energy_mesh}). The kinetic energy of the system is defined as:
\begin{equation}
    KE = \int_{\Omega} \frac{1}{2} \rho ||\boldsymbol{v}||^2 \mathrm{d}\Omega
    \label{eq:KE}
\end{equation}
and the strain energy for the incompressible neo-Hooekan solids is given by
\begin{equation}
    SE = \int_{\Omega_s} \frac{1}{2} \mu_s^L (tr(\boldsymbol{B})-d) \mathrm{d}\Omega
    \label{eq:SE}
\end{equation}
where $d$ is the dimension of the problem i.e. $2$ in our case. The verification of the above energy computations for an FSI case has been shown in \ref{appendixB}. We can observe that the kinetic energy is maximum at the initial time step due to the imposed velocity conditions and dissipates gradually because of the viscous fluid. The energy variations during this dissipation process are caused by the conversion of the kinetic energy into the solid strain energy and vice versa. The kinetic energy reaches its minimum when the solids are closest to each other. Concurrently, the solid strain energy peaks since the solids have achieved their maximum deformation. Eventually, the total energy decays owing to the viscosity of the fluid.

\subsubsection{Efficiency of MFIA}
\begin{figure}
    \begin{subfigure}[h]{0.48\textwidth}
        \centering
        \includegraphics[width=1\textwidth]{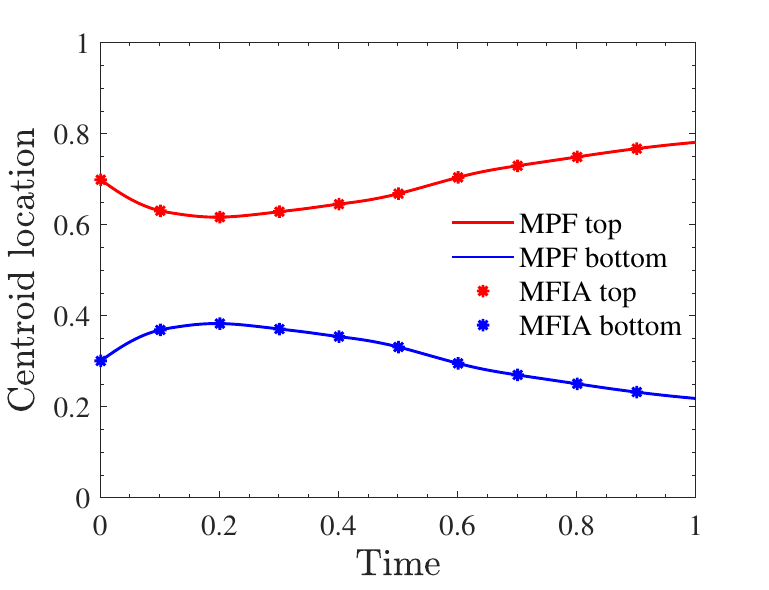}
        \caption{}
        \label{TG_traj_comp}
    \end{subfigure}
        \begin{subfigure}[h]{0.48\textwidth}
        \centering
        \includegraphics[width=1\textwidth]{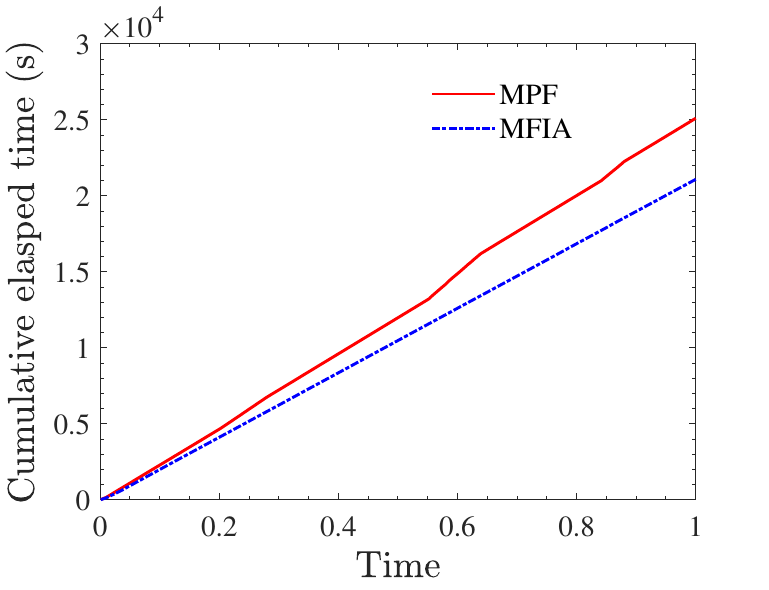}
        \caption{}
        \label{TG_time_comp}
    \end{subfigure}
    \caption{Comparison of (a) the centroid trajectories and (b) the total elapsed time between the two contact handling approaches (MFIA and MPF) for the two body collision problem.}
    \label{fig:TG_compare}
\end{figure}
In this section, we contrast the MFIA with the multi phase-field approach for detecting collision and resolving contact. For an FSI problem such as this example, it is not possible to have separate velocity fields for each solid body because of the presence of the background fluid. Thus, we have one momentum equation for the whole domain, two deformation equations, and three phase-field equations (two for the solid bodies and one for the background fluid). The implementation is similar to the multi phase-field approach carried out in \cite{mao20243d} or the multiple-level set methods in \cite{valkov2015eulerian, rycroft2020reference}. This approach essentially replaces the boundary update stage of the MFIA with additional PDEs nonetheless still requires the presence of a search algorithm to compute the distances during the broad phase collision detection step. We utilize the contact force subroutine as described for the MFIA.

Figure \ref{TG_traj_comp} shows the comparison of the centroid trajectories between the two algorithms to verify the results. To assess the cost effectiveness of the MFIA, we plot the total elapsed time for this test case using both approaches in Fig. \ref{TG_time_comp}. We observe a $16\%$ reduction in the total elapsed time for the MFIA as compared to the MPF approach. 
This can be attributed to the extra time consumed to solve the additional phase indicator and the strain equations in contrast to simply updating the solid boundaries. This gain is expected to become more prominent for greater number of solid bodies in the domain. It is worth mentioning that the MPF approach also results in increased memory requirements for the solver due to the introduction of the new solution variables that need to be saved after every time step.

\section{Applications} \label{sec7}
In this section, we present two applications to illustrate the robustness of the proposed phase-field based contact model. The first problem of a two-body collision system highlights an actuation mechanism for elastic bodies in fluid flow. The second test case demonstrates the capability of the solver to handle multibody contact problems involving complex hydrodynamic interactions.  

\subsection{Two Body Collision with Actuation}
Our approach also admits a simple method for solid actuation. This is especially advantageous when simulating clamped solids or prescribing a particular motion to the solid with applications ranging from active soft matter \cite{menzel2015tuned} to bio-locomotion studies \cite{trivedi2008soft, griffith2020immersed, bhosale2021remeshed}.

\subsubsection{Actuation of Bodies in Fluid Flow}
We utilize the constitutive model of the solid and the imposed kinematic conditions to achieve our objective. To actuate a particular solid, a smaller anchored region is defined inside the larger solid object. The strain tensor (i.e. the left Cauchy-Green deformation tensor $\boldsymbol{B}$ in the present case) is then set to be equal to the identity tensor $\boldsymbol{I}$ within the anchor zone. This restrains any deformation of the solid inside the anchor during the simulation. The next step is to specify the desired kinematic conditions in terms of imposed velocity in the anchored region. The rest of the body deforms in response to the imposed motion of the anchor. During the numerical implementation, we identify the nodes in the anchor as fixed nodes and remove them from the linear solve process. Another approach to achieve solid actuation can be to specify the desired deformation within the anchored zone as demonstrated in \cite{rycroft2020reference}. However, it is cumbersome to use this approach for cases where we do not know the exact deformation of the body or for cases where we would ideally want to impose a particular kinematic constraint.

\begin{figure}
    \centering
    \includegraphics[scale=0.65]{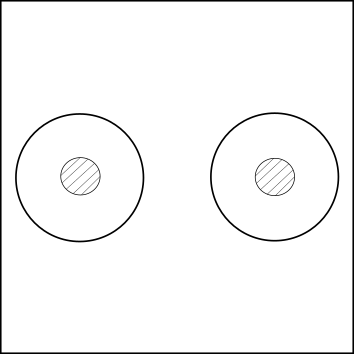}
    \put(-170,100){$\Omega_s^1$}
    \put(-50,110){$\Omega_s^2$}
    \put(-110,150){$\Omega_f$}
    \put(-158,70){\small{(fixed)}}
    \put(-63,70){\small{(moving)}}
    \caption{Schematic for the two body collision problem. The anchored regions have been shaded for both the bodies. The anchor on the left body is fixed whereas a harmonic x-velocity is provided to the anchor on the right body.}
    \label{fig:2body}
\end{figure}

\begin{figure}[htbp]
	\centering
	\begin{subfigure}{0.24\textwidth}
		\centering		
		\includegraphics[width=\textwidth]{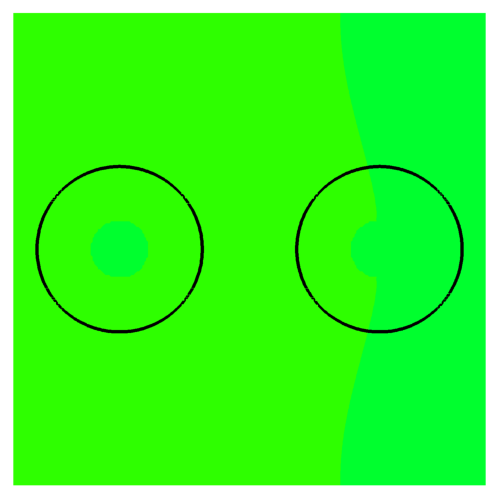}            
		\caption{}
	\end{subfigure}
	\hfill
	\begin{subfigure}{0.24\textwidth}
		\centering		
		\includegraphics[width=\textwidth]{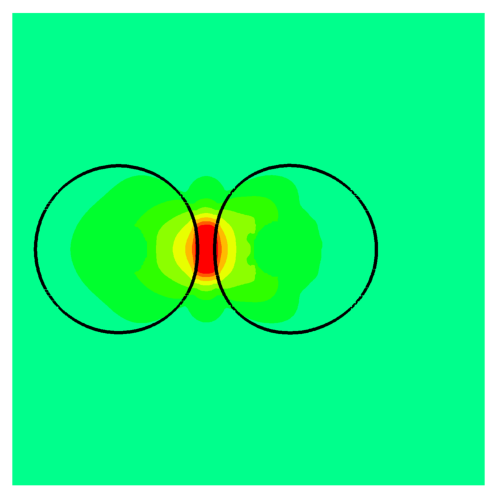}
		\caption{}
	\end{subfigure}
        \begin{subfigure}{0.24\textwidth}
		\centering		
		\includegraphics[width=\textwidth]{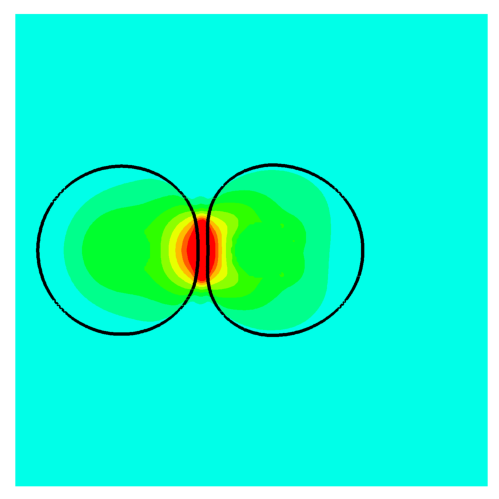}
		\caption{}
	\end{subfigure}
	\hfill
	\begin{subfigure}{0.24\textwidth}
		\centering		
		\includegraphics[width=\textwidth]{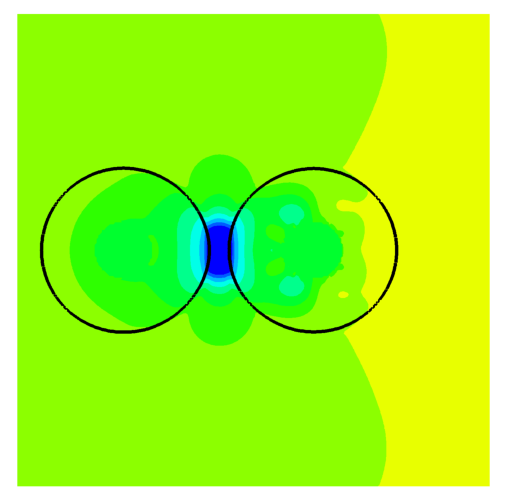}
		\caption{}
	\end{subfigure}

        \begin{subfigure}{0.48\textwidth}
		\centering		
		\includegraphics[width=\textwidth]{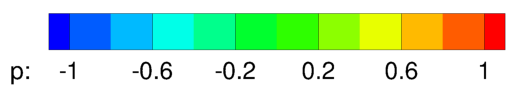}
		\label{legend_2body}
	\end{subfigure}
	\caption{Two body collision problem: Snapshots of the pressure contours at non-dimensional times $t=$ (a) $0$, (b) $0.36$, (c) $0.5$ and (d) $0.72$. The physical time has been non-dimensionalized with the time period of the imposed motion. The left ball is anchored at $(-1.1,0)$ and the right ball is given a prescribed velocity via an anchored region initially centred at $(1.1,0)$. The anchor for the right body gets updated as it moves through the domain.} The black solid lines denote the iso-contours with $\phi=0$ i.e. the boundaries of the two bodies.
	\label{fig:2body_collision}
\end{figure}

We present a test case to demonstrate the effectiveness of the proposed actuation approach. To illustrate the robustness of the unified approach, two incompressible hyperelastic solid bodies are considered that are actuated differently and undergo contact. This problem has also been carried out by \cite{valkov2015eulerian} in the context of an imposed reference/inverse map condition within a level set based finite difference discretization. The domain is a square of side length 4 on $[-2,2] \times [-2,2]$ and the radius of the circular discs are $R_1=R_2=0.7$ initially. The physical parameters for the problem are densities, $\rho_f=\rho_s=1$, fluid viscosity, $\mu_f=0.1$, shear modulus, $\mu_s^L=10$ and solid viscosity, $\mu_s=0.007$. The value of $\kappa$ is considered as 5 for this problem. Both bodies are provided with an anchored region of radius 0.25 at their center which controls their motion (Fig. \ref{fig:2body}). The left body is fixed at its initial location while the right body is given a harmonic motion of the following form:
\begin{align}
    v_x(t) = -\frac{\pi}{25}\sin \left(\frac{2\pi t}{25} \right).
\end{align}
We simulate this test case on a $256 \times 256$ grid with a time step of $0.005$. We normalize the physical time with the time period of the above-imposed motion, i.e. $T=25$. The initial condition for the order parameter in this problem is considered as follows:
\begin{equation}
    \begin{aligned}
        \phi(x,y,0) = 1 &+ \mathrm{tanh}\left( \frac{R_1 - \sqrt{(x+1.1)^2 + (y-0)^2}}{\sqrt{2}\varepsilon} \right) \\
    &+ \mathrm{tanh}\left( \frac{R_2 - \sqrt{(x-1.1)^2 + (y-0)^2}}{\sqrt{2}\varepsilon} \right).
    \end{aligned}
\end{equation}
The interface resolution is assumed to be $\frac{\varepsilon}{h}=2$.

Figure \ref{fig:2body_collision} illustrates the pressure contours obtained from the above implementation for two colliding bodies immersed in fluid. The results match qualitatively with those presented in \cite{valkov2015eulerian}. The snapshots confirm the accurate motion of the two bodies via the active solid actuation approach. We can observe that pressure builds up between the two bodies as they approach each other. The right body reaches its extremum at a non-dimensional time of $t=0.5$ and separates from the left body gradually due to the elastic deformations in the body and the application of the contact forces. We are able to avoid any numerical sticking between the objects via the implemented contact routine. A negative pressure region is created as it moves away and returns to its initial configuration.

\subsection{Free-falling of Elastic Bodies in Fluid}

\begin{figure}[htbp]
	\centering
	\begin{subfigure}{0.3\textwidth}
		\centering		
		\includegraphics[width=\textwidth]{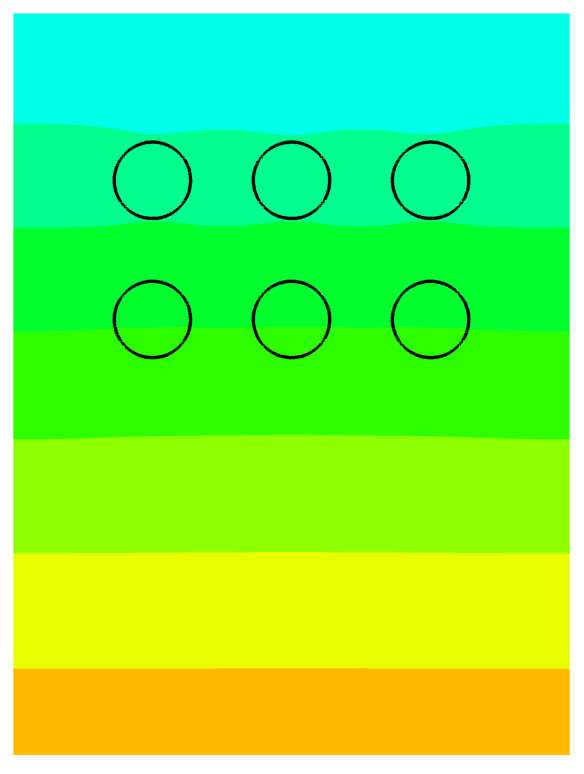}       
		\caption{}
	\end{subfigure}	
	\begin{subfigure}{0.3\textwidth}
		\centering		
		\includegraphics[width=\textwidth]{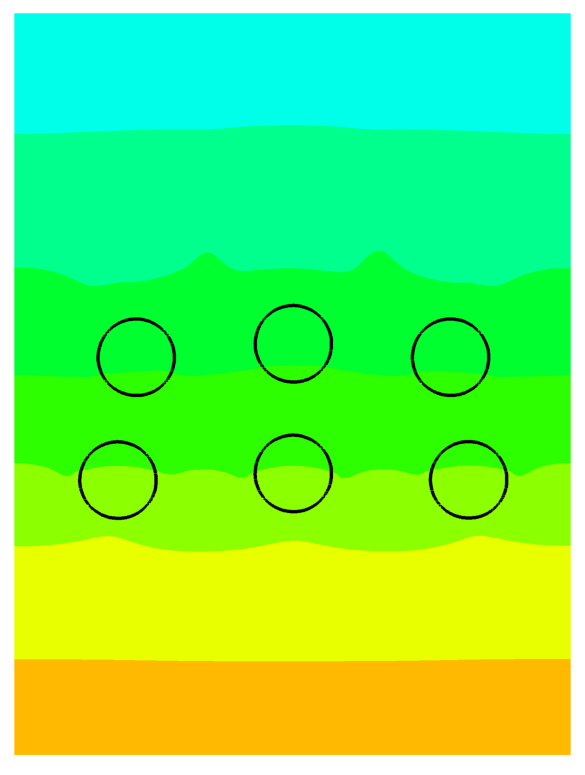}
		\caption{}
	\end{subfigure}
        \begin{subfigure}{0.3\textwidth}
		\centering		
		\includegraphics[width=\textwidth]{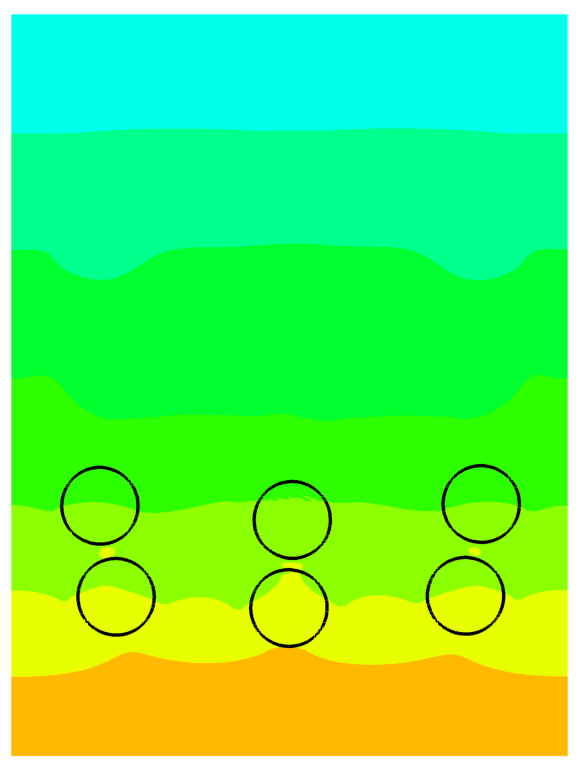}
		\caption{}
	\end{subfigure}	
	\begin{subfigure}{0.295\textwidth}
		\centering		
		\includegraphics[width=\textwidth]{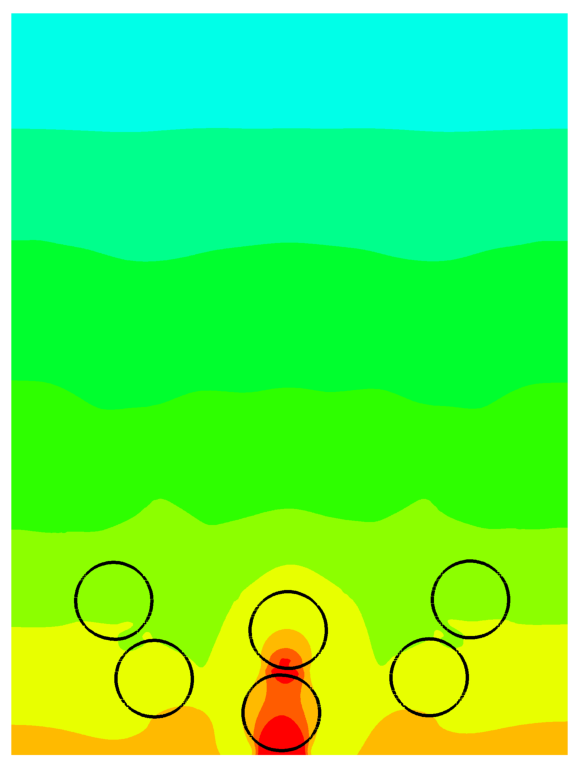}
		\caption{}
	\end{subfigure}        
	\begin{subfigure}{0.298\textwidth}
		\centering		
		\includegraphics[width=\textwidth]{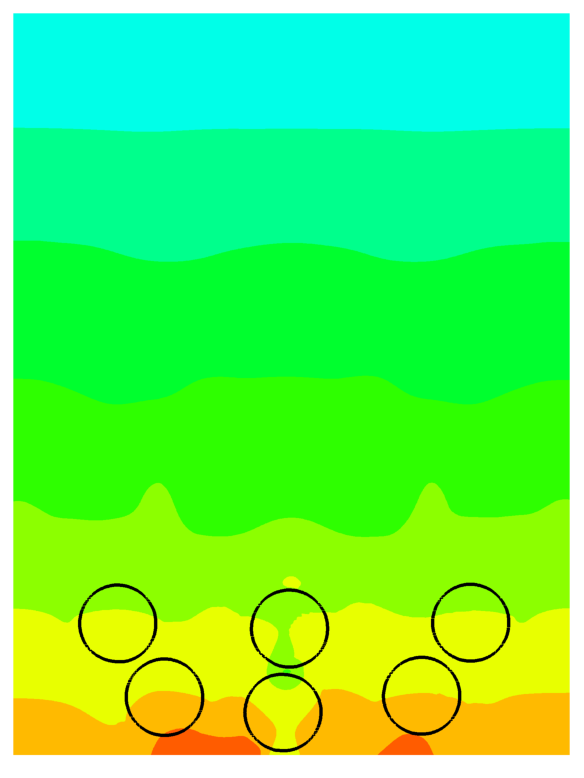}
		\caption{}
	\end{subfigure}        
	\begin{subfigure}{0.305\textwidth}
		\centering		
		\includegraphics[width=\textwidth]{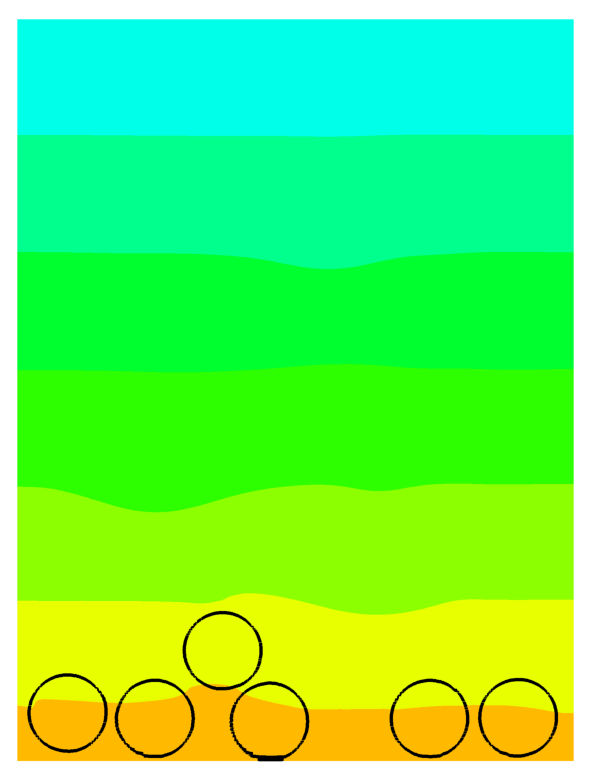}
		\caption{}
	\end{subfigure}

        \begin{subfigure}{0.35\textwidth}
		\centering		
		\includegraphics[width=\textwidth]{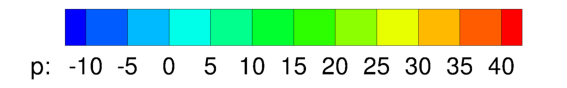}
		\label{legend_6body}
	\end{subfigure}
	\caption{Free-falling of elastic bodies: Snapshots of the pressure contours for the problem at $t=$ (a) $0$, (b) $3$, (c) $4.5$, (d) $5.5$, (c) $6$ and (f) $20$. The six bodies start at the top of the domain and fall freely under the action of gravity, undergoing complex hydrodynamic interactions and collision. The black solid lines denote the iso-contours with $\phi=0$ i.e. the boundaries of the bodies.}
	\label{fig:mb_pressure}
\end{figure}

In this section, we present a test case for demonstrating the robustness of the implemented multibody contact routine. We choose a sedimentation problem of six elastic bodies in an open tank \cite{liu2020nitsche}. The problem setup is similar to the suspension and sedimentation studies of particles carried out in \cite{glowinski2001fictitious, rakotonirina2018grains3d, reder2021phase} with applications in fluidized beds, sediment transport in rivers, and blood flow in arteries. The elastic solids are initialized at the top of the tank and allowed to fall freely under the action of gravity. The problem involves complex hydrodynamic interactions, elastic-elastic collisions between the solid bodies and elastic-rigid collisions between the bodies and the rigid walls. The domain is a rectangle on $[-3,3] \times [-7,1]$ and the radius of the circular discs are 0.4 initially. The physical properties of the problem are as follows: $\rho_f=1$, $\rho_s=2$, $\mu_f=0.005$ and $\mu_s^L=10^3$. A gravitational acceleration of $4$ units is applied in the downward direction. All physical parameters represent nondimensional values. The initial condition for the phase-field function is defined as:
\begin{equation}
    \phi(x,y,0) = 5 + \sum_{i=1}^6 \mathrm{tanh}\left( \frac{R_i - \sqrt{(x-x_C^i)^2 + (y-y_C^i)^2}}{\sqrt{2}\varepsilon} \right),
\end{equation}
where $(x_C^i,y_C^i)$ is the coordinate of the centroid and $R_i$ is the radius of the $i^{th}$ body. The interface resolution is considered to be $\frac{\varepsilon}{h}=2$ and a time step size of 0.005 is chosen. The value of $\kappa$ is assumed to be 0.5 for this problem. The no-slip boundary conditions for velocity are imposed on the left, right and bottom walls with the do-nothing boundary condition at the top to simulate an open tank. Homogeneous Neumann conditions are applied on all the boundaries of the Allen-Cahn equation.  

Figure \ref{fig:mb_pressure} illustrates the pressure contours for the above test case at six different time instants. The local pressure increases as the bodies approach the rigid walls or collide against each other. The applied contact forces ensure nonpenetration and allow the bodies to separate after colliding. 
After a long time, all the bodies settle at the bottom of the liquid tank (Fig. \ref{fig:mb_pressure}(f)).

\section{Conclusions} \label{sec8}
In the current work, we presented an efficient way of modeling multibody contact problems in a phase-field-based fully Eulerian framework for incompressible continua. Using a fixed mesh, our finite-element Eulerian framework can simulate large deformations and topology changes such as contact mechanics in a relatively simple manner. The phase-field method allows for a gradually increasing contact force implementation via its diffused interface representation. We verified our contact force model with the classical Hertz contact problem for smooth, dry contact between a cylinder and a plane. We also presented a comparison of force-indentation curves for longer simulations/ larger indentations with available analytical solutions. We assessed our model for a more general FSI contact scenario of two hyperelastic bodies colliding in a Taylor-Green vortex field against available literature. We showcased the volume conservation properties and the energy-transfer mechanisms for this problem via our Eulerian phase-field model. The relative volume error was shown to be within $5\%$ for all the cases simulated with this problem. The proposed MFIA method avoids the use of redundant interface capturing and strain evolution equations for multiple solids with identical physical properties. It also precludes the necessity of extracting the solid boundaries at every time step in case of diffused interface descriptions such as phase-field. These factors helped in reducing the computational time by around $16\%$ compared to a multi phase-field approach. Using our proposed phase-field Eulerian framework, we demonstrated two test cases, namely contact between two submerged bodies with actuation and the free-falling of multiple elastic bodies in an open tank.

The proposed framework can be extended to contact between compressible solids by restricting the divergence-free condition to the fluid domain exclusively and carrying out a harmonic extension of the fluid pressure field similar to \cite{dunne2006adaptive}. An alternative approach would be to use discontinuous finite elements to decouple fluid and structure pressure fields while satisfying the divergence condition in the fluid \cite{wick2013fully}. No additional considerations are required in the contact algorithm itself. It is worth noting that in the current implementation, we introduced an additional repulsive term in the momentum balance equation but did not alter the phase-field equation. This treatment works reasonably well for inertia-driven FSI applications similar to those considered herein. However, it might not be ideal for very long-time simulations of constant dry contact between solids as it can lead to unnecessary strain build-up. It might be worth exploring the inclusion of an additional energy potential in the Allen-Cahn equation to counteract the gradual diffusion of the phase field without constantly increasing the strain for such applications. In the present work, we have restricted ourselves to 2D problems for verification and demonstration purposes. The next step involves extending the present algorithm to 3D and integrating it with the multi phase-field solver \cite{mao20243d} to simulate the coupled ship-ice-water-air system. This will allow us to capture the rich multiphysics phenomena involving complex hydrodynamic and solid-solid interactions observed in a fully coupled ship-ice system.

\appendix
\setcounter{equation}{0}
\setcounter{figure}{0}

\section{Jacobian Computation for the Solid Stress Terms}
\label{appendixA}
In order to obtain the linearized form of the solid strain evolution, we utilize the generalized-$\alpha$ time integration scheme. As mentioned in Section \ref{sec2.3}, we decompose the solid stress into its volumetric and deviatoric components to obtain a uniform description of pressure. Hence, we will focus on deriving the Jacobian for the deviatoric component of a 2D stress state in this appendix.
We expand the deviatoric part of the solid stresses in terms of the left Cauchy-Green tensor, utilizing the incompressible neo-Hookean material model:
\begin{equation}
	\boldsymbol{\sigma}_{s}' = \mu^L_{s} (\boldsymbol{B}^{\mathrm{n}+\alpha} - \frac{1}{2} tr(\boldsymbol{B}^{\mathrm{n}+\alpha}) \boldsymbol{I}) .
\end{equation}
We need to evaluate $\boldsymbol{B}^{\mathrm{n}+\alpha}$ to substitute in the above equation to calculate the stresses. For this, we make use of the generalized-$\alpha$ time integration and the evolution equation for the left Cauchy-Green tensor. From the generalized-$\alpha$ time integration, we have
\begin{subequations}
\begin{align}
	\boldsymbol{B}^{\mathrm{n}+\alpha} &= \boldsymbol{B}^{\mathrm{n}} + \alpha(\boldsymbol{B}^{\mathrm{n}+1} - \boldsymbol{B}^{\mathrm{n}} ) ,\\
	&= \boldsymbol{B}^{\mathrm{n}} + \alpha \Delta t \left(\partial_t \boldsymbol{B}^{\mathrm{n}} + \frac{\varsigma}{\alpha_m} (\partial_t \boldsymbol{B}^{\mathrm{n+\alpha_m}} - \partial_t \boldsymbol{B}^{\mathrm{n}})\right) ,\\
	&= \boldsymbol{B}^{\mathrm{n}} + \alpha \Delta t \left( \left(1-\frac{\varsigma}{\alpha_m} \right)\partial_t \boldsymbol{B}^{\mathrm{n}} + \frac{\varsigma}{\alpha_m}\partial_t \boldsymbol{B}^{\mathrm{n+\alpha_m}}\right) , \label{B_n+alpha}
\end{align}
\label{gen_alpha_B}
\end{subequations}
where $\alpha$, $\alpha_m$ and $\zeta$ are the generalized-$\alpha$ parameters which are dependent on the user-defined spectral radius $\rho_{\infty}$:
\begin{align}
	\alpha = \frac{1}{1+\rho_{\infty}}, \text{   } \alpha_m = \frac{1}{2} \left( \frac{3-\rho_{\infty}}{1+\rho_{\infty}} \right), \text{   } \zeta = \frac{1}{2} - \alpha + \alpha_m .
\end{align}
Using the transport equation for the left Cauchy-Green tensor, we have
\begin{equation}
	\partial_t \boldsymbol{B}^{\mathrm{n+\alpha_m}} = \nabla \boldsymbol{v}^{\mathrm{n+\alpha}}\boldsymbol{B}^{\mathrm{n+\alpha}} + \boldsymbol{B}^{\mathrm{n+\alpha}} (\nabla \boldsymbol{v}^{\mathrm{n+\alpha}})^T -(\boldsymbol{v}^{\mathrm{n+\alpha}} \cdot \nabla)\boldsymbol{B}^{\mathrm{n+\alpha}} .
\label{B_transport}
\end{equation}
For simplification, we assume $\varsigma=\alpha_m$ i.e. $\rho_{\infty}=1$. Substituting Equation (\ref{B_transport}) in Equation (\ref{B_n+alpha}) with the above simplification, we obtain
\begin{equation}
	\boldsymbol{B}^{\mathrm{n}+\alpha} = \boldsymbol{B}^{\mathrm{n}} + \alpha \Delta t \bigg(\nabla \boldsymbol{v}^{\mathrm{n+\alpha}}\boldsymbol{B}^{\mathrm{n+\alpha}} + \boldsymbol{B}^{\mathrm{n+\alpha}} (\nabla \boldsymbol{v}^{\mathrm{n+\alpha}})^T -(\boldsymbol{v}^{\mathrm{n+\alpha}} \cdot \nabla)\boldsymbol{B}^{\mathrm{n+\alpha}} \bigg) .
\end{equation}
The Jacobian or the tangent stiffness matrix can thus be evaluated as:
\begin{equation}
    \boldsymbol{K} = \frac{\partial \boldsymbol{\sigma}_s'}{\partial \boldsymbol{v}^{\mathrm{n}+\alpha}} = \mu_s^L \left( \frac{\partial \boldsymbol{B}^{\mathrm{n}+\alpha}}{\partial \boldsymbol{v}^{\mathrm{n}+\alpha}} - \frac{1}{2} \frac{\partial (tr(\boldsymbol{B}^{\mathrm{n}+\alpha})\boldsymbol{I})}{\partial \boldsymbol{v}^{\mathrm{n}+\alpha}}  \right).
\end{equation}
The terms of the above equation can be evaluated as shown in \cite{mao2021variational}.

\section{Oscillating Disk in Taylor-Green Vortex Field}
\label{appendixB}
\begin{figure}[htbp]
	\centering
	\begin{subfigure}{0.24\textwidth}
		\centering		
		\includegraphics[width=\textwidth]{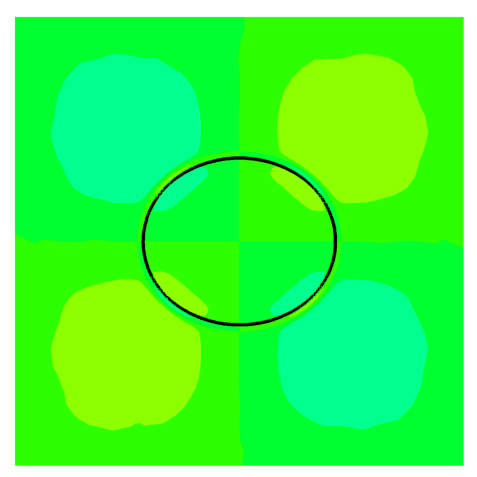}            
		\caption{}
	\end{subfigure}
	\hfill
	\begin{subfigure}{0.24\textwidth}
		\centering		
		\includegraphics[width=\textwidth]{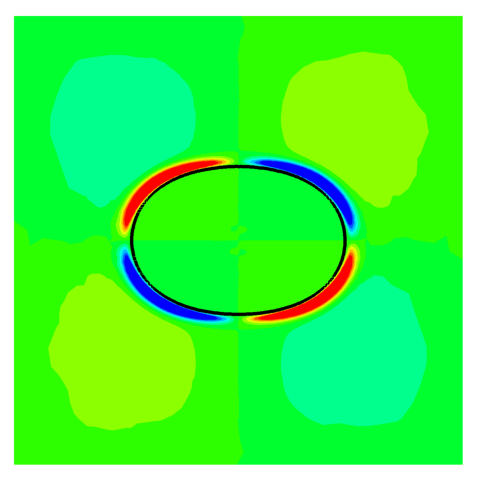}
		\caption{}
	\end{subfigure}
        \begin{subfigure}{0.24\textwidth}
		\centering		
		\includegraphics[width=\textwidth]{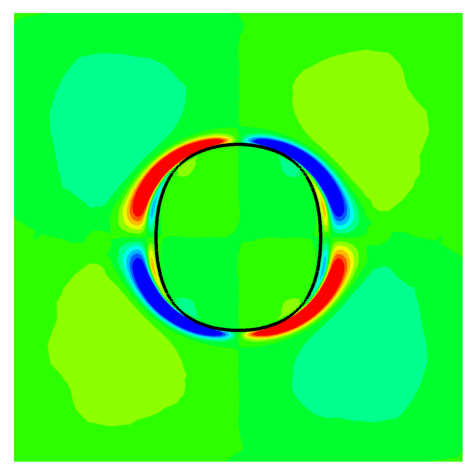}
		\caption{}
	\end{subfigure}
	\hfill
	\begin{subfigure}{0.24\textwidth}
		\centering		
		\includegraphics[width=\textwidth]{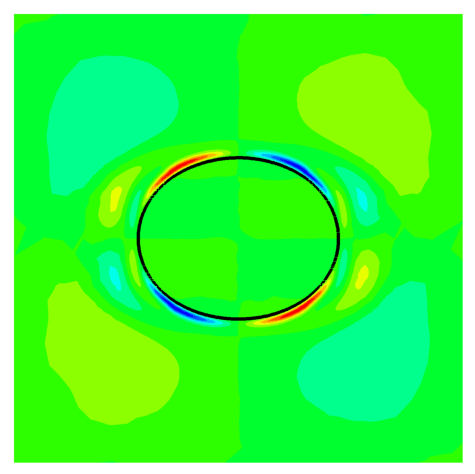}
		\caption{}
	\end{subfigure}

    \begin{subfigure}{0.35\textwidth}
		\centering		
		\includegraphics[width=\textwidth]{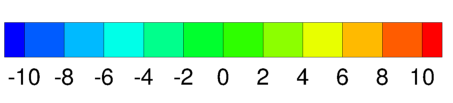}
        \put(-185,6){$\omega_z:$}
		\label{legend_TG_1body}
	\end{subfigure}   
	\caption{Oscillating disk in a Taylor-Green voretx field: Snapshots of the vorticity contours for the oscillating disk problem at $t=$ (a) $0$, (b) $0.25$, (c) $0.5$ and (d) $1$. The black solid line denotes the iso-contour with $\phi=0$ for the disk.}
	\label{fig:TG_1body}
\end{figure}

\begin{figure}
    \centering
    \includegraphics[scale=0.95]{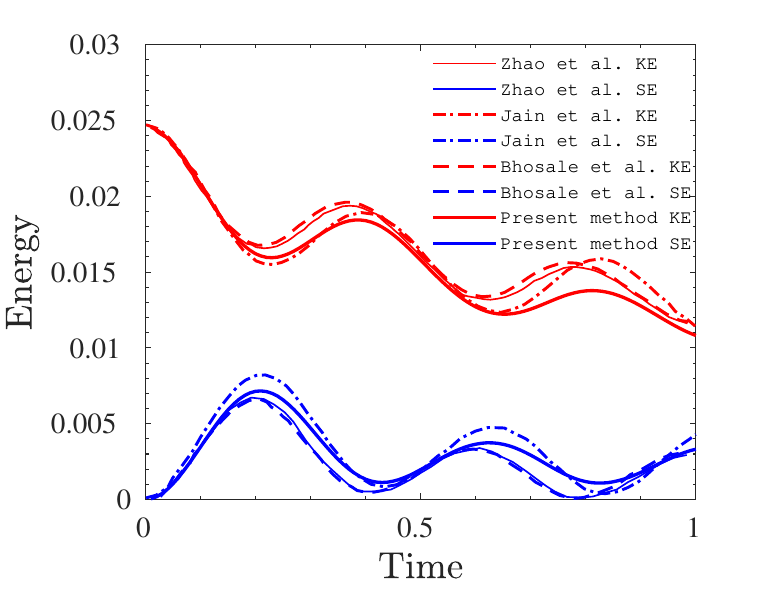}
    \caption{Comparison of temporal variation of kinetic energy KE and strain energy SE with previous works \cite{zhao2008fixed, jain2019conservative, bhosale2021remeshed}.}
    \label{fig:TG_1energy}
\end{figure}
We present our numerical results for the FSI case of an oscillating disk placed in a Taylor-Green vortex field and compare it with available literature to verify our energy computations. A unit square box over $[0,1]\times [0,1]$ with periodic boundaries is defined as the computational domain for this problem. An incompressible hyperelastic solid is placed at the center of the domain ($0.5,0.5$) at $t=0$. The stress-free undeformed configuration of the solid is a circle of radius $R=0.2$. The initial condition for the phase-field function is considered as follows:
\begin{equation}
    \phi(x,y,0) = 1 + \mathrm{tanh} \left( \frac{R - \sqrt{(x-0.5)^2 + (y-0.5)^2}}{\sqrt{2}\varepsilon}  \right).
\end{equation}
The interface resolution is chosen $\frac{\varepsilon}{h}=1$. A Taylor-Green vortex field, similar to Section \ref{TG_2body}, is initialized with the following streamfunction distribution:
\begin{equation}
    \psi(x,y) = \psi_0 \mathrm{sin}(k_xx) \mathrm{sin}(k_yy),
\end{equation}
where $\psi_0=0.05$ and $k_x=k_y=2\pi$. Other physical parameters for the simulation are fluid viscosity $\mu_f=10^{-3}$, shear modulus $\mu_s^L=1$ and density $\rho_f=\rho_s=1$. A structured grid of $100\times 100$ and a time step size of $0.001$ have been used for this problem. 

Figure \ref{fig:TG_1body} illustrates the vorticity contours for this problem in different time steps. The dynamics of the solid body resemble that of a damped oscillator. The solid is initially deformed by the imposed vorticity and retracts as a result of its elastic response. This generates oscillations in the system that dissipate eventually due to the viscosity of the fluid.
We track the temporal evolution of the kinetic energy of the system and strain energy of the solid as defined in Eqs. \ref{eq:KE} and \ref{eq:SE}. Figure \ref{fig:TG_1energy} shows the comparison of the energy plots obtained from our approach with previous works in the literature. This verifies our energy calculations and highlights that our results are consistent with a variety of other methods.


\bibliography{ref.bib}

\end{document}